\documentclass[aps,prx,twocolumn,footinbib]{revtex4-2}
\usepackage{graphicx}
\usepackage{indentfirst}
\usepackage{physics}
\usepackage{braket}
\usepackage{float}
\usepackage{amsmath}
\usepackage{epstopdf}
\usepackage{CJK}
\usepackage{esint}
\usepackage{color}
\usepackage[T1]{fontenc}
\usepackage{subfigure}
\usepackage{amsfonts}
\usepackage{footmisc}
\usepackage{scrextend}
\usepackage{multirow}
\usepackage{amssymb}
\usepackage[hyperfootnotes=false]{hyperref}
\usepackage{multirow}
\usepackage[english]{babel}
\usepackage{url}
\usepackage{bm}
\usepackage{hyperref}
\definecolor{darkblue}{rgb}{0,0,0.5}
\hypersetup{
colorlinks=true,
linkcolor=black,
filecolor=blue,
citecolor=darkblue,  
urlcolor=black,
}

\newcommand*\diff{\mathop{}\!\mathrm{d}}

\newtheorem{theorem}{Theorem}

\newtheorem{lemma}[theorem]{Lemma}

\newcommand\argmin{\mathop{\mathrm{argmin}}}
\newcommand{\calA}{{\cal A}}

\newcommand{\calL}{{\cal L}}
\newcommand{\calN}{{\cal N}}

\newcommand{\calU}{{\cal U}}

\newcommand{\1}{^{(1)}}

\def\be{\begin{equation}}
\def\ee{\end{equation}}
\def\ba{\begin{eqnarray}}
\def\ea{\end{eqnarray}}
\usepackage{bm}

\newcommand{\QZ}[1]{{{\textcolor{blue}{#1}}}}

\begin{document}

\title{Continuous-variable error correction for general Gaussian noises}

\author{Jing Wu$^{1}$}
\author{Quntao Zhuang$^{1,2}$}
\email{zhuangquntao@email.arizona.edu}

\address{
$^1$James C. Wyant College of Optical Sciences, University of Arizona, Tucson, AZ 85721, USA
}
\address{
$^2$Department of Electrical and Computer Engineering, University of Arizona, Tucson, Arizona 85721, USA
}

\begin{abstract}
Quantum error correction is essential for robust quantum information processing with noisy devices. As bosonic quantum systems play a crucial role in quantum sensing, communication, and computation, it is important to design error correction codes suitable for these systems against various different types of noises. While most efforts aim at protecting qubits encoded into the infinite dimensional Hilbert space of a bosonic mode, [Phys. Rev. Lett. 125, 080503 (2020)] proposed an error correction code to maintain the infinite-dimensional-Hilbert-space nature of bosonic systems by encoding a single bosonic mode into multiple bosonic modes. Enabled by Gottesman-Kitaev-Preskill states as ancilla, the code overcomes the no-go theorem of Gaussian error correction. In this work, we generalize the error correction code to the scenario with general correlated and heterogeneous Gaussian noises, including memory effects. We introduce Gaussian pre-processing and post-processing to convert the general noise model to an independent but heterogeneous collection of additive white Gaussian noise channels and then apply concatenated codes in an optimized manner. To evaluate the performance, we develop a theory framework to enable the efficient calculation of the noise standard deviation after the error correction, despite the non-Gaussian nature of the codes. Our code provides the optimal scaling of the residue noise standard deviation with the number of modes and can be widely applied to distributed sensor-networks, network  communication and composite quantum memory systems. 
\end{abstract} 


\maketitle

\section{Introduction}
Quantum information science has brought to us novel capabilities in computing~\cite{Shor_1997}, sensing~\cite{caves1981quantum} and communication~\cite{Bennett20147}. Continuous-variable (CV) systems play an essential role in quantum information science, due to its deterministic quantum resource generation and robustness against noises. CV measurement-based quantum computing~\cite{menicucci2006universal,menicucci2014fault} is one of the promising route towards universal quantum computing, and CV cluster states have been experimentally demonstrated in a large scale~\cite{asavanant2019generation}. Super-conducting qubits~\cite{blais2020quantum}, with demonstrated benefit from error correction~\cite{ofek2016extending}, are based on the CV modes of a cavity-QED system. Equipped with quantum resources including squeezing and entanglement, CV systems have also enabled quantum advantages in many sensing applications, including phase sensing~\cite{escher2011general}, loss sensing~\cite{nair_2018}, distributed sensing~\cite{zhuang2018distributed,zhang2020distributed}, spectroscopy~\cite{shi2020PRL}, quantum target detection~\cite{Tan2008,zhuang2017entanglement,pirandola2011quantum,zhuang2020entanglement-enhanced}, and most prominently the Laser Interferometer Gravitational-Wave Observatory~\cite{LIGO,LIGO_nat,haocun2020quantum}. Besides computing and sensing, communication channels such as fiber or atmosphere links can be modeled as bosonic channels, which naturally requires a CV description. Indeed, CV encoding is important in classical communication~\cite{giovannetti2014ultimate}, entanglement-assisted classical communication~\cite{Bennett2002,shi2020,zhuang2020entanglement} and quantum key distribution~\cite{grosshans2002,zhuang2016,pirandola2015high}.


To enable these quantum advantages in an experimental setting, quantum error correction (QEC)~\cite{calderbank1996} is the rosetta stone to reduce degradation from noise and loss. For computing purposes, the pioneering work of Gottesman, Kitaev and Preskill (GKP)~\cite{gottesman2001} provides a method of encoding qubits into oscillators (CV modes), whose code states are later named the GKP states and have recently been experimentally engineered~\cite{fluhmann2019encoding,campagne2020quantum}. Utilizing GKP states, CV measurement-based quantum computation is made fault-tolerant~\cite{menicucci2014fault,baragiola2019}. The CV nature of the errors also lead to advantages in the discrete encoding of quantum information~\cite{fukui2017}. As the ideal GKP states have infinite energy, effects from approximate versions of GKP states have also been analyzed~\cite{tzitrin2020}.

QEC has also enabled enhanced performance in quantum sensing applications~\cite{zhou2018achieving,layden2019ancilla,zhuang2020distributed}, where CV systems are ubiquitous. To protect quantum information in CV systems, recent progress~\cite{noh2019encoding} has provided a method of encoding an oscillator into multiple oscillators, utilizing entangling Gaussian operations on GKP ancilla. 
As a first step, Ref.~\cite{noh2019encoding} addresses an independent and identical (iid) collection of additive white Gaussian noise (AWGN) channels and provides GKP-Gaussian codes to reduce the noise on the data mode. However, dynamical fluctuations and memory effects in quantum systems can often lead to heterogeneous and correlated noises, including scenarios related to distributed sensing, quantum network communication and quantum memory systems (see Fig.~\ref{schematic_concept}). In fact, practical systems are always finitely-correlated and two quantum systems will never precisely encounter {\it exactly} identically-distributed noises. Moreover, such correlated errors over a large length scale will often make usual QEC schemes impotent. Therefore, to enable CV QEC's advantage, the capability to correct general forms of noises is crucial.

In this paper, we address the general Gaussian noise model, where the noises between different modes can be heterogeneous and correlated. We show that a general multi-mode noisy Gaussian channel~\cite{Weedbrook_2012} can be reduced to a product of independent but heterogeneous AWGN channels. Then we formulate a general GKP-Gaussian error correction scheme design and obtain a lower bound of the reduced logical noise. We provide a framework to enable the efficient evaluation of the performance of different codes, including a multi-mode concatenation of GKP-two-mode-squeezing (TMS) code and the GKP-squeezing-repetition (SR) code. Our numerical results show that tailoring the error correction procedure optimally for each noise model will provide a huge advantage in the noise-reducing capability. A global optimization enables the concatenated codes to reach the optimal scaling versus the input noise level; while greedy optimization leads to a worse scaling in general. Finally, we apply our machinery on a Gaussian memory channel~\cite{lupo2010capacities,caruso2014quantum} to demonstrate the QEC performance. We also demonstrate the fidelity improvement from the QEC in protecting a single-mode squeezed-vacuum state.

This paper is organized as follows: in Section~\ref{sec:noise_model}, we introduce the general noise model and reduction to the standard form; in Section~\ref{sec:general_QEC},  we design the general GKP-Gaussian error correction scheme and derive its ultimate lower bound in the output noise standard deviation (STD); Section~\ref{sec:two_mode} provides performance evaluations of the two-mode case, while the multi-mode case is solved in Section~\ref{sec:multi_mode}. As an example, we apply the code on the Gaussian memory channel in Section~\ref{sec:memory_channel}. Finally, we conclude with some discussions in Section~\ref{sec:conclusion}.

\section{Noise model: correlated Gaussian noise and memory effects}
\label{sec:noise_model}

\begin{figure}
\centering
\includegraphics[width=0.45\textwidth]{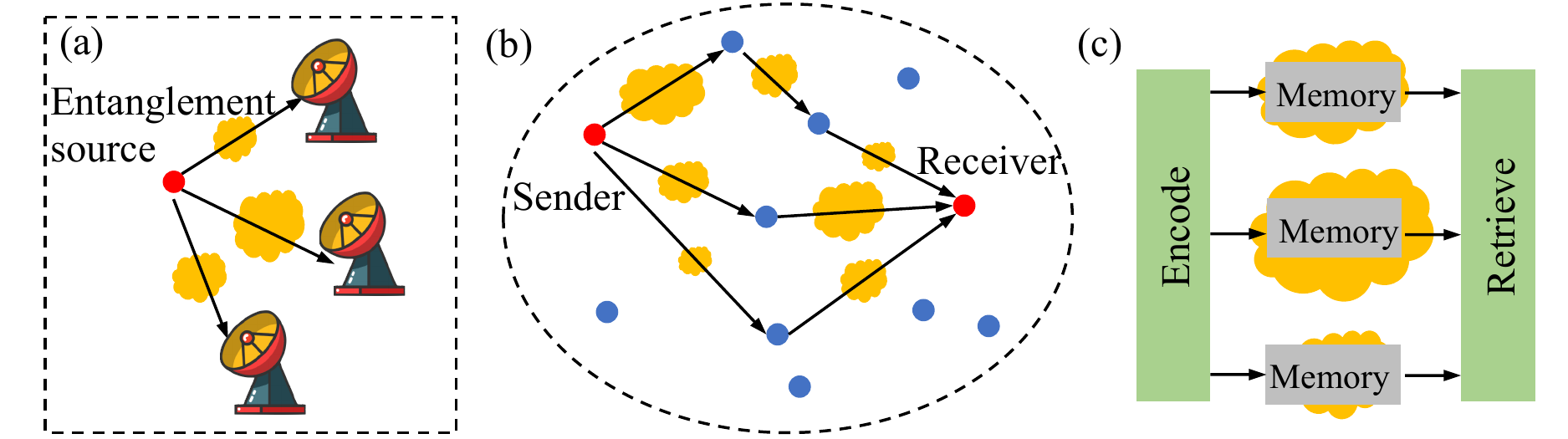}
\caption{Concept of scenarios with heterogeneous and/or correlated noises. 
(a) In distributed sensing scenarios, different sensors can operate in very different environments~\cite{zhuang2018distributed,zhuang2020distributed}; (b) In a quantum network~\cite{kimble2008quantum,wehner2018quantum,kozlowski2019towards,zhang2019}, when one utilizes multi-path routing~\cite{pirandola2019end}, the noise levels between different paths are going to be heterogeneous; (c) Even in well-controlled quantum memories, the noises can defer among different parts. The different size of the yellow `cloud' indicates the heterogeneous noise structure.
\label{schematic_concept}
}
\end{figure}

A common imperfection in a CV system can be modelled as a thermal-loss channel $\calL_{\eta,N_B}$, as is the case for applications summarized in Fig.~\ref{schematic_concept}. 
The channel $\calL_{\eta,N_B}$ can be described by the beamsplitter transform over the annihilation operators.
\be 
\hat{a}\to \sqrt{\eta} \hat{a}+\sqrt{1-\eta}\hat{e}_L,
\ee 
where $\hat{a}$ and $\hat{e}_L$ represent the input mode and  the environmental mode separately. Here $\eta$ is the transmissivity of the loss channel and the environment mode $\hat{e}_L$ is in a thermal state with the mean photon number $\braket{\hat{e}_L^\dagger \hat{e}_L}=N_B/(1-\eta)$, where we have chosen to set the thermal photon mixed into the output as $N_B$.

In the phase space, such a thermal-loss channel has two effects on the Wigner function of the input state: the $\sqrt{\eta}$ factor shrinks the coordinates of the Wigner function; while the noise term $N_B$ increases the variances. To enable error correction against a thermal-loss channel, one can first make use of a channel-concatenation relation to first eliminate the shrinking effects. Namely, one applies an amplifier $\calA_{G}$ with the mode transform
\be 
\hat{a}\to \sqrt{G} \hat{a}+\sqrt{G-1}\hat{e}_G^\dagger,
\ee 
on the input prior to the thermal-loss channel, where $\hat{e}_G$ is in a vacuum state. Choosing $G=1/\eta$, we have the overall channel
\be 
\calL_{\eta,N_B} \circ \calA_{1/\eta} = \Phi_{N_B+1-\eta},
\label{single_mode-reduction}
\ee 
as an AWGN channel with variance $N_B+1-\eta$. Then Ref~\cite{noh2019encoding} shows that such an AWGN noise can be error-corrected, utilizing Gaussian unitary and GKP ancilla, as we explain in Section~\ref{sec:general_QEC}.

\begin{figure*}
\centering
\includegraphics[width=0.75\textwidth]{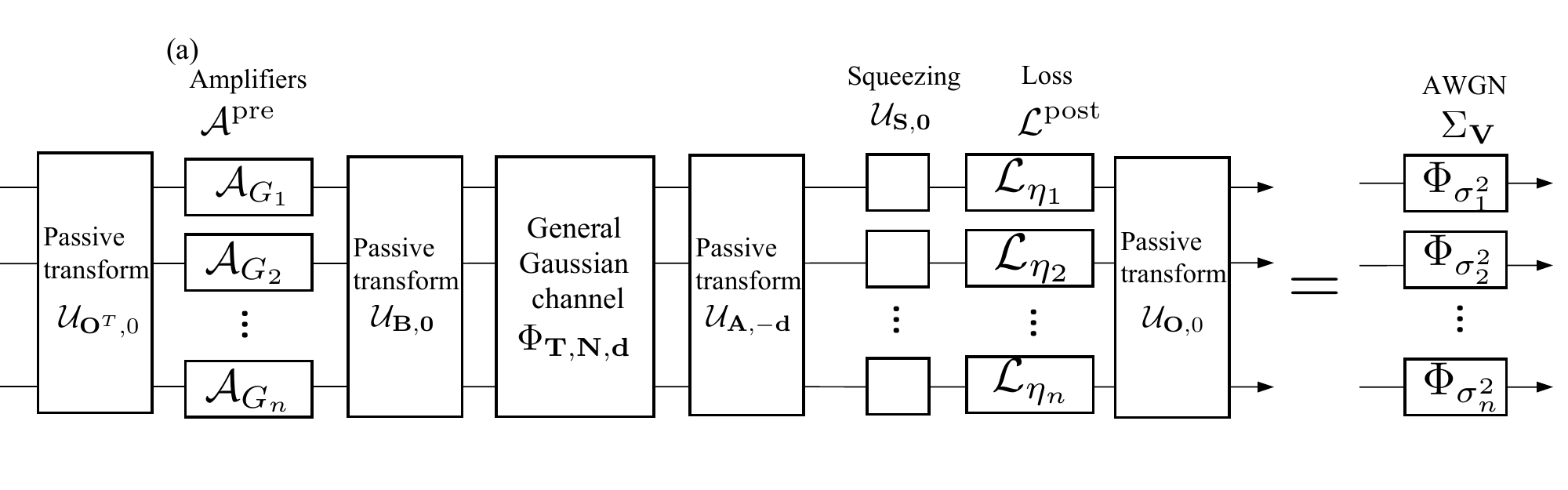}
\caption{Schematic of the general Gaussian channel reduction scheme to a product of additive white Gaussian noise (AWGN) channels.
\label{schematic_reduction}
}
\end{figure*}


However, real applications often require multiple modes, as depicted in Fig.~\ref{schematic_concept}. When multiple modes are involved, the noise model can be more complicated. For example, in quantum communication, one would utilize multiple time or frequency modes to transmit the overall code-word. Due to the temporal or spectral fluctuations in the transmission links, the noises among different modes can be heterogeneous. Moreover, the noises can be correlated due to frequency cross-talk or memory effects. In this regard, an applicable noise model is an $n$-mode general Gaussian channel~\cite{Weedbrook_2012,Holevo2001} $\Phi_{\bm T, \bm N, \bm d}$, which in general mixes the input modes, introduces squeezing, amplifies or attenuates the $n$ input modes and introduces additional noises. Here $\bm d$ is a $2n$-dimensional vector of displacement, $\bm T$ is a $2n\times 2n$ real matrix to characterize the mixing and squeezing, and $\bm N$ is a $2n\times 2n$ real matrix to characterize the additive noises. Despite the complexity, we show the following lemma to map the noise model to a standard collection of AWGN channels. 
\begin{lemma}
\label{lemma:general_reduction}
Any multi-mode Gaussian channel can be transformed to a product of independent AWGN channels, i.e.,
\be 
\Sigma_{\bm V}=\bigotimes_{\ell=1}^n \Phi_{\sigma_\ell^2},
\label{noise_model}
\ee 
via Gaussian pre-processing and post-processing. Here the matrix $\bm V={\rm Diag}\left[\sigma_1^2,\cdots,\sigma_n^2\right]$ is the covariance matrix of the AWGN noises.
\end{lemma}

The full proof is presented in Appendix~\ref{app:proof_reduction}, here we provide the intuition of the reduction process. As shown in Fig.~\ref{schematic_reduction}, we can introduce two passive unitary transforms $\calU_{\bm B,0}$ and $\calU_{\bm A, -\bm d}$ as pre- and post-processing to cancel the mixing between different modes; then we can introduce individual squeezers in the post-processing to adjust the squeezing in each mode; afterwards, the pre-amplification $\calA^{\rm pre}$ and post-attenuation (loss) $\calL^{\rm post}$ adjust the shrinking effects to map the overall channel to a correlated AWGN channel, similar to the single-mode pure loss case in Eq.~\eqref{single_mode-reduction}; finally, a conjugation of two passive unitary operations $\calU_{\bm O^T,0}$ and $\calU_{\bm O,0}$ disentangles the correlation between the noise to obtain the final independent AWGN channel in Eq.~\eqref{noise_model}. We will consider an explicit example of a Gaussian memory channel in Sec.~\ref{sec:memory_channel}.

To provide intuition on the QEC design, we introduce the random displacements interpretation of an AWGN channel. First, we introduce the position and momentum quadrature operators: $\hat{q}=\left(\hat{a}^\dagger+\hat{a}\right)/\sqrt{2}$ and $\hat{p}=i\left(\hat{a}^\dagger-\hat{a}\right)/\sqrt{2}$, with the canonical communication relation $[\hat{q},\hat{p}]=i$. Under this definition, the vacuum noise $\braket{\hat{p}^2}=\braket{\hat{q}^2}=1/2$. Formally, the action of an AWGN channel with an STD $\sigma$ on the single-mode input state $\hat{\rho}$ can be described by
\be 
\Phi_{\sigma^2} \left(\hat{\rho}\right)=\int d^2\bm x F_{\sigma}\left(x_1\right) F_{\sigma}\left(x_2\right) \hat{D}\left(x_1,x_2\right)\hat{\rho} \hat{D}^\dagger\left(x_1,x_2\right),
\label{phi_N_disp}
\ee 
where $F_{\sigma}\left(\cdot\right)$ is a zero-mean Gaussian probability density function (PDF) with the STD $\sigma$. Here the displacement operator $\hat{D}\left(x_1,x_2\right)=\exp\left[-i \left(x_1\hat{p}-x_2 \hat{q}\right)\right]$ acting on the input state effectively shifts the input quadratures by $\hat{q}\to \hat{q}+x_1, \hat{p}\to \hat{p}+x_2$. 

\begin{figure}[b]
\centering
\includegraphics[width=0.45\textwidth]{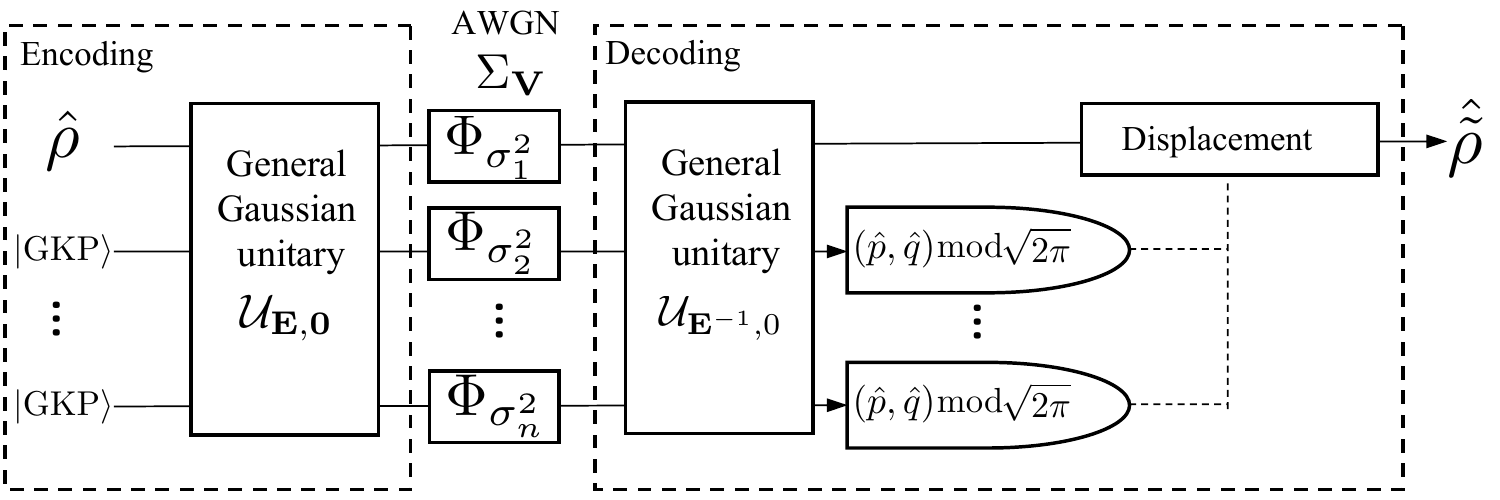}
\caption{
The general multi-mode GKP-Gaussian code. The notation $\calU$ denotes the channel that a unitary transform $\hat{U}$ incurs.
\label{fig:general_code}
}
\end{figure}

\section{General GKP-Gaussian error correction and the ultimate lower bound}
\label{sec:general_QEC}
To correct the AWGN noises, Ref.~\cite{noh2019encoding} adopts an approach of combining GKP grid states and Gaussian operations, to go beyond the no-go theorem for Gaussian QEC~\cite{niset2009nogo}. There, an iid noise model is adopted to examplify the basic principle; Here, we present a general framework that extends Ref.~\cite{noh2019encoding}'s approach to the multi-mode general AWGN channel in Eq.~\eqref{noise_model}. As depicted in Fig.~\ref{fig:general_code}, the QEC scheme applies a general zero-displacement Gaussian unitary $\hat{U}_{\bm E,\bm 0}$, described by the symplectic transform $\bm E$, on the input state $\hat{\rho}$ and $n-1$ GKP ancilla to obtain an entangled $n$-mode non-Gaussian state. After passing through the AWGN channel $\Sigma_{\bm V}$, the output goes through the inverse of the Gaussian unitary $\hat{U}_{\bm E^{-1},\bm 0}$. Finally, one simultaneously measures the momentum and position quadratures (up to module $\sqrt{2\pi}$) of the $n-1$ GKP ancilla modes via GKP-assisted measurements. Based on the measurement results, one applies a displacement on the data mode to obtain an approximation of the original input $\hat{\tilde{\rho}}$.

The intuition behind the QEC design is that the channel concatenation,
\be 
\calU_{\bm E^{-1},\bm 0}\circ \Sigma_{\bm V} \circ \calU_{\bm E,\bm 0}=\Sigma_{\bm E^{-1}\bm V (\bm E^{-1})^T},
\ee 
correlates the errors on different modes in a controlled way. As an AWGN can be interpreted as random displacements via Eq.~\eqref{phi_N_disp}, measuring the displacements on the $n-1$ ancilla will lead to a good estimation of the displacement error remaining on the data mode. The crucial contribution of the GKP grid state is to enable the joint measurement of displacements on both quadratures, as shown in Refs.~\cite{terhal2016encoding,duivenvoorden2017}, up to a module $\sqrt{2\pi}$ ambiguity, so that the joint estimation of both quadrature displacements is possible. This can be confirmed by the wave function of a GKP grid state
\begin{align}
&\ket{{\rm GKP}}\propto 
\sum_{t=-\infty}^\infty e^{-\pi \Delta^2 t^2 }
\int e^{-(q-\sqrt{2\pi}t)^2/2\Delta^2} \ket{q} dq
\nonumber
\\
&\propto 
\sum_{t=-\infty}^\infty  \int e^{-\Delta^2p^2/2} e^{-(p-\sqrt{2\pi}t)^2/2\Delta^2}\ket{p}dp,
\end{align} 
where $\ket{p}$ and $\ket{q}$ are the momentum and position eigenstates. 
When $\Delta\ll1$, its Wigner function is peaked around a square grid of spacing $\sqrt{2\pi}$ in the phase space. The overall variance $\braket{\hat{q}^2}\simeq \braket{\hat{p}^2}\simeq 1/2\Delta^2$ equals the mean photon number $N_S$; however, if we consider only the phase space region close to a single peak, the variances in position and momentum are $\Delta^2/2\simeq 1/4N_S\ll 1$, only twice the squeezed-vacuum variance. In this paper, we will consider the $\Delta\to 0$ limit of ideal GKP states.

In the above scheme, we have delayed all measurement till the end; in general, one can consider performing measurements and displacements in between the Gaussian gates. However, due to the Gaussian nature of the unitaries, the displacements conditioned on the measurement results can always be pushed until the end, and even largely avoided via post-processing on the measurement results. Indeed, only the displacements on the data mode is necessary in general.

To judge the performance of a QEC scheme, as AWGN noise applies random displacements, we can consider the STDs of the random displacement errors after the decoding procedure. In general, the STD $\sigma_{L,q}$ on the position quadrature and the STD $\sigma_{L,p}$ on the momentum quadrature can be unequal, therefore we define the average logical noise STD $\sigma_L=\sqrt{(\sigma_{L,q}^2+\sigma_{L,p}^2)/2}$.
Regardless of the error correction scheme, for a noise model described in Eq.~\eqref{noise_model}, one can prove a lower bound of the noise STD
\be 
\sigma_L^2\ge \frac{1}{e} \prod_{\ell=1}^n \frac{\sigma_\ell^2}{1-\sigma_\ell^2}>  \frac{1}{e} \prod_{\ell=1}^n \sigma_\ell^2,
\label{sigma_L_limit}
\ee 
which generalizes Eq. (12) of Ref.~\cite{noh2019encoding} to the multi-mode case (see Appendix~\ref{proof_lower_bound} for a proof). Note we will only consider $0<\sigma_\ell<1$, as the case of zero noise is trivial and channels with $\sigma_\ell>1$ will be discarded due to zero quantum capacity.

Below, we consider two specific families of codes, the GKP-TMS code and GKP-SR code and compare their performances with the ultimate limit in Eq.~\eqref{sigma_L_limit}.

\section{Performance with a single GKP ancilla}
\label{sec:two_mode}

We begin with the two-mode case ($n=2$) of the noise model in Eq.~\eqref{noise_model}, where the noise covariance $\bm V={\rm Diag}(\sigma_1^2,\sigma_1^2,\sigma_2^2,\sigma_2^2)$ for a data mode and a single GKP ancilla. 

\subsection{GKP-two-mode squeezing code}
In a GKP-TMS-code, the symplectic transform of the encoding
\begin{align}
    \bm E = 
    \begin{pmatrix}
    \sqrt{G} \boldsymbol{I} && \sqrt{G-1} \boldsymbol{Z}\\
    \sqrt{G-1} \boldsymbol{Z} && \sqrt{G} \boldsymbol{I}\\
    \end{pmatrix}
\end{align}
corresponds to a TMS operation, where $\boldsymbol{I} = {\rm Diag}(1,1)$ and $\boldsymbol{Z} = {\rm Diag}(1,-1)$. The random displacement $\boldsymbol{z} = (z_q^{(1)},z_p^{(1)},z_q^{(2)},z_p^{(2)})$ of the output AWGN channel now has the covariance matrix
\begin{align}
    \label{covMatrix2}
    \begin{split}
    &\boldsymbol{V}_{\boldsymbol{z}} = (\bm E)^{-1} \boldsymbol{V} (\bm E^{-1})^{\boldsymbol{T}}=\\
    &
    \begin{pmatrix}
        [G\sigma_1^2+(G-1)\sigma_2^2]\boldsymbol{I} && -\sqrt{G(G-1)}(\sigma_1^2+\sigma_2^2)\boldsymbol{Z}\\
        -\sqrt{G(G-1)}(\sigma_1^2+\sigma_2^2)\boldsymbol{Z} && [G\sigma_2^2+(G-1)\sigma_1^2]\boldsymbol{I}
    \end{pmatrix}.
    \end{split}
\end{align}
Below we summarize the decoding procedure, with full details presented in Appendix~\ref{App:TMS_code}.

As mentioned, we can measure $z_q^{(2)} $ and $z_p^{(2)}$ module $\sqrt{2 \pi}$, with the outcome ($\tilde{z}_q^{(2)}=R_{\sqrt{2\pi}}(z_q^{(2)})$, $\tilde{z}_p^{(2)}=R_{\sqrt{2\pi}}(z_p^{(2)})$), where $R_{s}(z)$ stands for generalized module of $s$, i.e., $R_s(z)=z-n^\star(z)s$ and $n^\star(z)=\argmin_{n\in \mathbb{Z}}|z-ns|$. We denote the set of all integers $\mathbb{Z}$. 
Let’s first assume that the results $(\tilde{z}_q^{(2)}$, $\tilde{z}_p^{(2)})$ are exactly $(z_q^{(2)}$, $z_p^{(2)})$, then we adopt the minimum variance estimator for the unknown displacement error on the first mode $\bar{z}_x^{(1)} = H_x\tilde{\mu}\tilde{z}_x^{(2)}$ for $x=p,q$. The constant $\tilde{\mu}={\sqrt{G(G-1)}(\sigma_1^2+\sigma_2^2)}/{\sigma_G^2}$, where we have introduced $\sigma_G^2=(G-1)\sigma_1^2+G\sigma_2^2$, and the signs $H_p=1, H_q=-1$. With the estimator in hand, one performs the displacement $\hat{D}\left(-\bar{z}_q^{(1)},-\bar{z}_p^{(1)}\right)$, attempting to cancel the noise on the data mode. 

\begin{figure}
\centering
\includegraphics[width=0.475\textwidth]{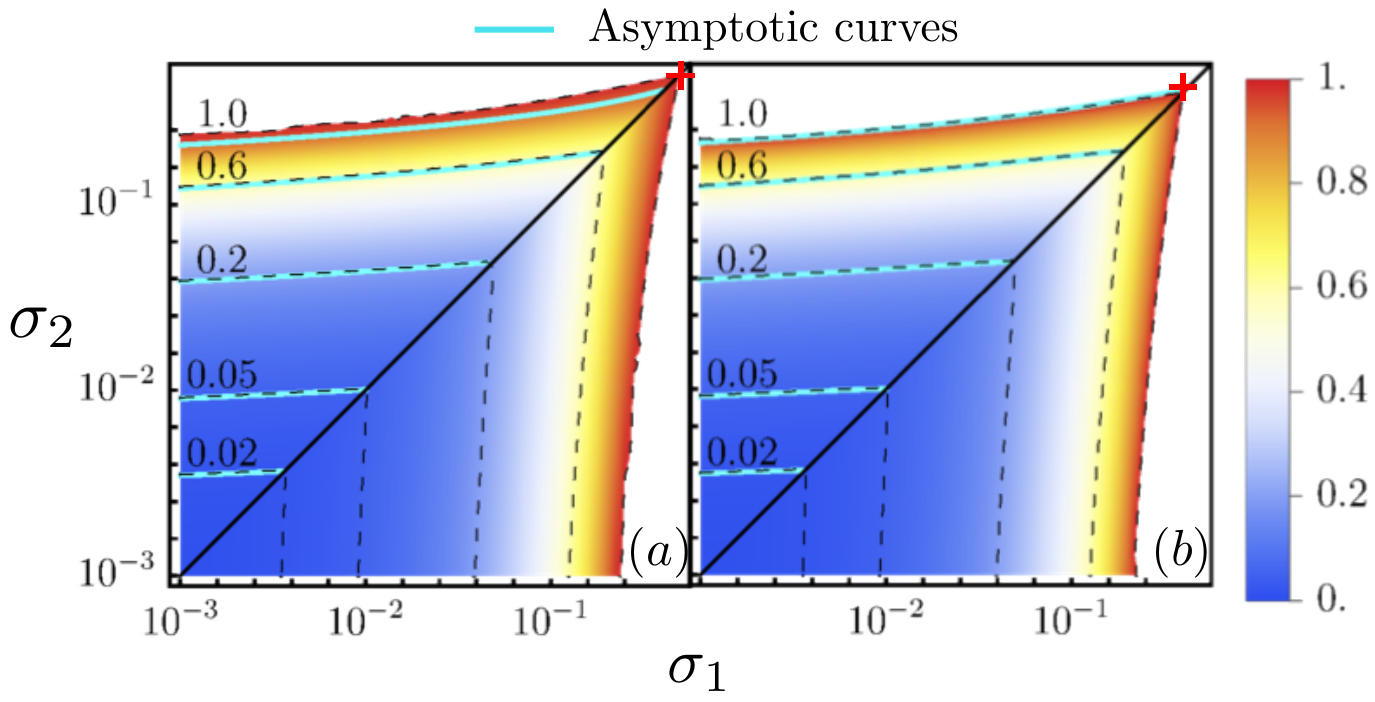}
\caption{ Contours of the error-correction ratio $\sigma_L^\star/\min[\sigma_1,\sigma_2]$ for 
(a) GKP-two-mode squeezing code and
(b) GKP-squeezing-repetition code.
The end points marked by the red crosses are $\sigma_1=\sigma_2\simeq 0.56$ for (a) and $\sigma_1=\sigma_2\simeq 0.41$ for (b).
\label{fig:contours}
}
\end{figure}

In the following, we analyze the residual noise of the above error correction procedure.
To begin with, the PDFs $Q^{(2)}(\cdot)$ and $P^{(2)}(\cdot)$ of the residue quadrature noises $\xi_q=z_q^{(1)}-\bar{z}_q^{(1)}$ and $\xi_p=z_p^{(1)}-\bar{z}_p^{(1)}$ can be obtained through the joint Gaussian distribution function (see Appendix~\ref{App:TMS_code} for details). 
It turns out that
\begin{align}
&Q^{(2)}(x)=P^{(2)}(x)=f(x;\sigma_G,\sigma^{(2)},\tilde{\mu})
\nonumber
\\
& \equiv \sum_{n\in\mathbb{Z}}b_n(\sigma_G) \; F_{\sigma^{(2)}}(x+\tilde{\mu}\sqrt{2\pi}n),
\label{sum_f}
\end{align}
where $\sigma^{(2)}={\sigma_1\sigma_2}/{\sigma_G}$ and the function 
\be 
\label{eq:exact_bn}
b_n(\sigma)
\equiv \frac{1}{2}\left\{{\rm Erfc}\left[\frac{(n-\frac{1}{2})\sqrt{\pi}}{\sigma}\right]
-{\rm Erfc}\left[\frac{(n+\frac{1}{2})\sqrt{\pi}}{\sigma}\right]\right\}.
\ee 
With the PDFs in hand, we can proceed to obtain the identical output noise variance as
\begin{align}
\label{eq:exact_sigma}
\sigma_L^2=g(\sigma_G, \sigma^{(2)}, \tilde{\mu}) \equiv \left(\sigma^{(2)}\right)^2 + \sum_{n\in\mathbb{Z}} b_n(\sigma_G)\tilde{\mu}^2 2\pi n^2,
\end{align}
which can be numerically evaluated efficiently.

\begin{figure*}
\centering
\includegraphics[width=0.95\textwidth]{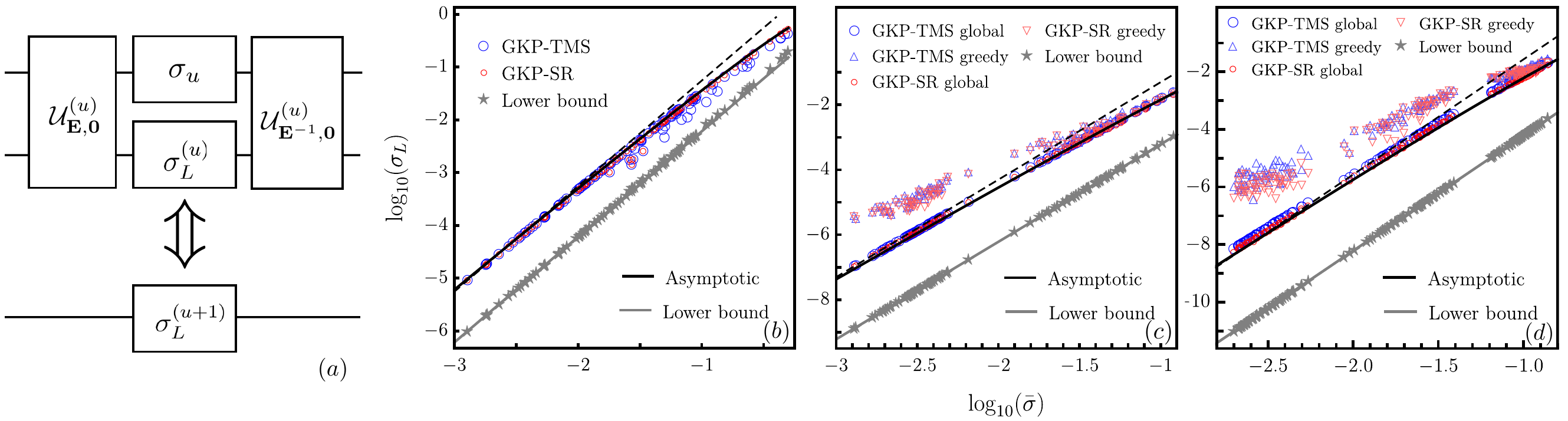}
\caption{(a) Schematic of the concatenation of codes. (b-d) The corrected noise STD $\sigma_{L}^\star$ vs $\bar{\sigma}$ in a logarithmic scale. The number of modes equals two (b), three (c) and four (d). Asymptotic results (black solid lines) are obtained from Eq.~\eqref{eq:asymptotic_u} for comparison. The black dashed lines show the scaling of $\sigma_L^{\star}\sim \bar{\sigma}^n$, with $n=2,3,4$ in (b)(c)(d). The lower bound (gray solid lines and stars) comes from Ineq.~\eqref{sigma_L_limit}.  
\label{fig:sigma_scaling}
}  
\end{figure*}

To minimize the above output noise variance, we can further optimize over the gain $G\ge1$---a large gain creates a good correlation between the noises to enable error correction, while introduces extra noises due to the module $\sqrt{2\pi}$ uncertain. When both $\sigma_1$ and $\sigma_2$ are small, we can obtain the asymptotic result of the optimal noise variance
\begin{align}
\label{eq:asympotic expressions_TMS}
{\sigma_L^\star}^2 = \upsilon (\sigma_1^2,\sigma_2^2)\approx
\frac{4 \bar{\sigma}^4}{\pi}\ln\left(\frac{\pi^{3/2}}{2\bar{\sigma}^4}\right),
\end{align}
where the geometric mean $\bar{\sigma}=\left(\sigma_1\sigma_2\right)^{1/2}$.
When $\sigma_1=\sigma_2$, the above result agrees with the asymptotic result of Ref.~\cite{noh2019encoding}. While Eq.~\eqref{eq:asympotic expressions_TMS} is symmetric between $\sigma_1$ and $\sigma_2$, the exact result is not---numerical results show that the variance of the output noise is smaller when the smaller noise acts on the data mode, which can also be shown from the next-order analyses (see Eq.~\eqref{AppendixC11}). Therefore, we switch the order of the channels whenever $\sigma_1 > \sigma_2$ to enable a better performance with $\sigma_1 \le \sigma_2$. Fig.~\ref{fig:contours}(a) shows the contour plots of the ratio of the minimum output noise over the input noise, $\sigma_L^\star/\min(\sigma_1,\sigma_2)$. We see the ratio of noise reduction is better when both noises are smaller. As shown, the asymptotic approximation (cyan lines) in Eq.~\eqref{eq:asympotic expressions_TMS} also fits well with the numerical calculations for small $\sigma_1$ and $\sigma_2$.

To further verify our asymptotic result in Eq.~\eqref{eq:asympotic expressions_TMS}, we apply the numerical analyses to random samples of $(\sigma_1,\sigma_2)$ and plot the optimal corrected noise STD $\sigma_L^\star$ versus the geometric mean $\bar{\sigma}$. The samples have $\log_{10}\left(\sigma_1\right)$ and $\log_{10}\left(\sigma_2\right)$ uniformly distributed in the range of $[-3,-0.3]$. In Fig.~\ref{fig:sigma_scaling}(b), we see that the numerical results (open blue circles) agree well with the asymptotic Eq.~\eqref{eq:asympotic expressions_TMS} (black solid line), especially when $\bar{\sigma}$ is small. On the other hand, the corrected logical noise level $\sigma_L$ has the same scaling as the ultimate limit in Eq.~\eqref{sigma_L_limit} (gray line for the further lower bound and gray stars for more accurate lower bounds for each sample), but overall around one order of magnitude larger.


\subsection{Squeezing-repetition code}

In a GKP-SR code, the encoding Gaussian unitary
is composed of three single-mode squeezing operations and a two-mode SUM gate (see Appendix~\ref{App:sqz_rep} for details), leading to the overall symplectic transform
\begin{align}
\label{eq:symplectic transform of sq-rep}
\bm E &=\begin{pmatrix}
{\kappa}/{G} && 0 && 0 && 0\\
0 && {G}/{\kappa} && 0 && -G\\
G && 0 && {G}/{\kappa} && 0\\
0 && 0 && 0 && {\kappa}/{G}
\end{pmatrix},
\end{align}
which is parameterized by $\kappa$ and $G$. The decoding procedure is similar to the GKP-TMS code, where a minimum variance estimator is devised and a displacement operation is applied on the data mode accordingly, leading to the output noise PDFs similar to Eq.~\eqref{sum_f} as
\begin{subequations}
\begin{align}
&Q^{(2)}(x)=f(x;G \frac{\sigma_2}{\kappa}, \frac{\kappa \sigma_1}{G},\kappa \frac{\sigma_1}{\sigma_2}),
\\
&P^{(2)}(x)=f(x; G \frac{\sigma_2}{\kappa}, \frac{\kappa \sigma_1}{G}, \kappa),
\end{align}
\label{QP_distribution_SR}
\end{subequations}
where the function $f$ is defined in Eq.~\eqref{sum_f} and 
\be 
\kappa^2 = \left({\sqrt{G^8 \sigma_1^4+ 4 G^4{\sigma_2}^4}-G^4\sigma_1^2}\right)/{2\sigma_2^2}
\ee 
is chosen to balance the variances of each Gaussian peak in the noise PDF. Note that when $G\to0$, we have $\bm E\to\bm I$ approaches the identity, which corresponds to no error-correction. The variances of the output noise after error-correction can be obtained as
\begin{align}
\sigma_{L,q}^2 =g(G \frac{\sigma_2}{\kappa},\frac{\kappa \sigma_1}{G},\kappa \frac{\sigma_1}{\sigma_2}),
\sigma_{L,p}^2 =g(G \frac{\sigma_2}{\kappa}, \frac{\kappa \sigma_1}{G}, \kappa),
\end{align}
where the function $g$ is defined in Eq.~\eqref{eq:exact_sigma}.

Similar to the GKP-TMS code, here we optimize the parameter $G$ to obtain the minimum average noise $\sigma_L^\star$. When both $\sigma_1$ and $\sigma_2$ are small, asymptotically we have
\be 
\label{eq:asympotic expressions_SR}
{\sigma_L^\star}^2 \approx \frac{4\bar{\sigma}^4}{\pi} \ln \left[\frac{\pi^{3/2}}{{2\bar{\sigma}^4}}\right]+\frac{4\bar{\sigma}^4}{\pi} \ln\left(\frac{\sigma_1^2+\sigma_2^2}{2\sigma_1^2}\right),
\ee 
which is identical to Eq.~\eqref{eq:asympotic expressions_TMS} in the leading-order, up to a next-order correction that disappears when $\sigma_1=\sigma_2$. When $\sigma_1\neq \sigma_2$, while to the leading-order the GKP-TMS code is symmetric between the two channels, the GKP-SR code is asymmetric: Eq.~\eqref{eq:asympotic expressions_SR} shows that we will choose the order $\sigma_2\le \sigma_1$ to minimize $\sigma_L^\star$---we want to use the less noisy channel for GKP ancilla, in contrary to the GKP-TMS code.

We plot the contour of the ratio of the optimal average noise STD $\sigma_L^\star$ versus the minimum of the input STDs $\sigma_1,\sigma_2$ in Fig.~\ref{fig:contours}(b). The asymptotic results (cyan curves) agree well with the numerical results. We also perform the numerical analyses for the same set of random samples in the GKP-TMS case; as shown in Fig.~\ref{fig:sigma_scaling} (b) by the red open circles, the performance is similar to the GKP-TMS code in most cases. As the leading order asymptotic result is identical to Eq.~\eqref{eq:asympotic expressions_TMS}, we also see a good agreement with the asymptotic black solid curve.

\section{Multi-mode concatenation}
\label{sec:multi_mode}

A common technique in QEC code design is concatenation---each element in a single error-correction circuit layer can be further error-corrected by another layer of circuit, and thereby the noise is further suppressed. As the number of concatenation layers increases, the logical noise can usually be reduced to an arbitrary small amount. Fig.~\ref{fig:sigma_scaling}(a) shows the $(u+1)$-th layer of encoding, where a two-mode code is applied on the logical mode at the $u$-th layer (with noise STD $\sigma^{(u)}_L$) and another GKP ancilla mode (with noise STD $\sigma_u$). Thereby, one can further reduce the average noise STD to $\sigma^{(u+1)}_L$. For example, in the initial layer ($u=2$), the mode with noise STD $\sigma^{(2)}_L\equiv \sigma_1$ is encoded with a mode with noise STD $\sigma_2$; after the error-correction, the reduced noise has a variance $\left(\sigma^{(3)}_L\right)^2\simeq \upsilon(\sigma_1^2,\sigma_2^2)$ to the leading-order, as given in Eq.~\eqref{eq:asympotic expressions_TMS}.

Naively, one can simply use Eq.~\eqref{eq:asympotic expressions_TMS} recursively to obtain the leading-order logical noise in multiple layers of error-correction. For example, for the three-mode case of $u=4$ with $\sigma_1,\sigma_2$ at the bottom layer, we have
\begin{align}
    &{\sigma^{(4)}_L}^2 \simeq \upsilon \left(\sigma_3^2,{\sigma^{(3)}_L}^2\right)\simeq 
    \upsilon \left(\sigma_3^2,\upsilon\left(\sigma_2^2,\sigma_1^2\right)\right)
    \\
    &=\frac{16\bar{\sigma}^6}{\pi^2}
    \ln{\left(\frac{\pi^{3/2}}{2\sigma_1^2\sigma_2^2}\right)} \ln{\left[\frac{\pi^{5/2}}{8\bar{\sigma}^6}/\ln{\left(\frac{\pi^{3/2}}{2\sigma_1^2\sigma_2^2}\right)}\right]},
\end{align}
which is not symmetric between $\sigma_\ell$'s. Therefore, an optimization over the different orders of the noisy channels in the encoding is necessary, as we will explain later. To obtain a qualitative understanding, we consider a rough estimation using the geometric mean $\bar{\sigma}$ (see details in Appendices~\ref{App:TMS_concatenation} and~\ref{App:sqz_rep}),
\be 
{\sigma^{(u)}_L}^2\simeq \upsilon({\sigma_{u-1}^2},{\sigma^{(u-1)}_L}^2), u\ge 3,
\label{eq:asymptotic_u}
\ee 
which is exact to the leading-order for the special case of $u=3$. In Fig.~\ref{fig:sigma_scaling}, Eq.~\eqref{eq:asymptotic_u} is shown as the solid black curves for comparison. One can also show the scaling $\sigma^{(u)}\sim \bar{\sigma}^{u-1}$ when the noise STDs are small.

However, things are more complicated due to the non-Gaussian noise PDF. In the two-mode case, the initial displacement noise PDF is Gaussian, and becomes non-Gaussian as expressed in Eq.~\eqref{sum_f} after a single layer of error-correction. Therefore, a more accurate analysis is necessary to evaluate the performance of a concatenated code design.
Formally, for the noise model described by Eq.~\eqref{noise_model}, an $n$-layer concatenation of the two-mode codes is described by two vectors. First, the set of parameters $\bm G=(G_1,\cdots, G_{n-1})$ determines the encoding operations. Second, a permutation vector $\bm \Pi=(P_1,\cdots,P_n)$ denotes the order of channels $\{\Phi_{\sigma_\ell^2}\}_{\ell=1}^n$ being utilized: at the $u$-th layer, the additional mode goes through the channel $\Phi_{\sigma_{P_u}^2}$. Without loss of generality, we consider the noises in the channels to be ordered from small to large as $\sigma_1 \le \sigma_2 \cdots \le \sigma_n$. 
For example, the reverse order $\bm \Pi=(n,n-1,\cdots,3,2,1)$ describes the following procedure: the mode at the bottom layer experiences the the least noisy channel $\Phi_{\sigma_1^2}$, and then after the encoding with all other modes layer by layer, interacts with the data mode that goes through the most noisy channel $\Phi_{\sigma_n^2}$ (as indicated by the case of Fig.~\ref{fig:sigma_scaling}(a)).

With the encoding specified above, in both the GKP-TMS code and the GKP-SR code, after performing the displacement based on estimators similar to the two-mode case, it turns out that the residue noise PDF is a symmetric sum of Gaussian functions. For the GKP-TMS code, the PDFs of both quadratures at the $(u+1)$-th layer are identical,
\begin{equation}
P(\xi^\prime) = \sum_{k,\ell\in\mathbb{Z}}b_{k,\ell}F_{\sigma^{(u+1)}}\left(\xi^\prime-t_{k,\ell}\right),
\label{P_xi_general}
\end{equation}
where the parameters $b_{k,\ell}$, $\sigma^{(u+1)}$ and $t_{k,\ell}$ depend on parameters $\bm G$ and $\bm \Pi$. Therefore, we can keep track of the noise PDF efficiently, via a recursion (see Appendix~\ref{App:TMS_concatenation}). The GKP-SR code has a similar noise PDF composed of a sum of Gaussian distributions, however, asymmetric between the two quadratures (see Appendix~\ref{App:sqz_rep}).

\begin{figure}
\centering
\includegraphics[width=0.475\textwidth]{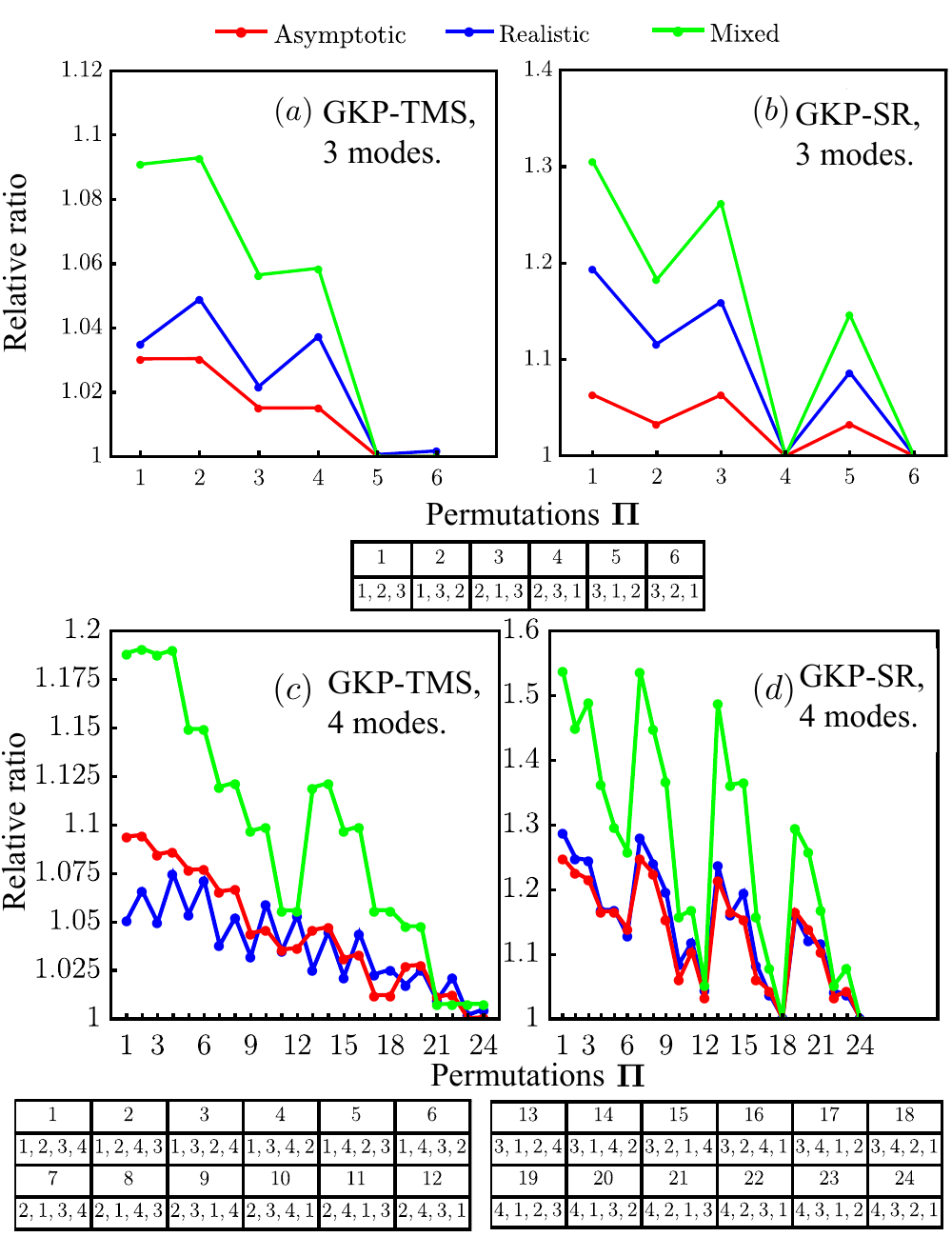}
\caption{Average relative ratios $\sigma_{L}^\star(\bm \Pi)/\min_{\bm \Pi} \sigma_{L}^\star(\bm \Pi)$ versus permutation $\bm \Pi$ for GKP-TMS and GKP-SR codes with three and four modes. See Fig.~\ref{fig:PermutationList_TMS} and Fig.~\ref{fig:PermutationList_SqzRep} in the Appendix for more details.
\label{fig:permutations_3_4}
}  
\end{figure}

For a concatenated scheme over the noise model in Eq.~\eqref{noise_model}, we can choose the parameter $\bm G$ and the order $\bm \Pi$ of the $n$ channels to minimize the output noise STD
\begin{equation}
    \sigma_{L}\left(\bm G, \bm \Pi\right) = \sqrt{\left(\sigma^{(n+1)}\right)^2 + \sum_{k,\ell\in\mathbb{Z}} b_{k,\ell} t_{k,\ell}^2}
\end{equation}
for the GKP-TMS code
or a similar formula for the GKP-SR code.
For $n$ channels, there are $n!$ choices of orders. For each order $\bm \Pi$, we may optimize the parameter $\bm G$ through a multi-parameter global optimization to obtain $\sigma_{L}^\star(\bm \Pi)=\min_{\bm G}\sigma_{L}\left(\bm G, \bm \Pi\right)$. Alternatively, one can adopt a greedy strategy: at the $u$-th layer ($3\le u \le n+1$) one minimizes $\sigma^{(u)}$ over a single parameter $G_{u-2}$ at this layer. We denote the greedy minimum as $\sigma_{L}^{\circledast}(\bm \Pi)$. Overall, a global scheme involves a joint optimization problem of $n-1$ parameters; while the greedy scheme only involves a single-parameter optimizing problem for $n-1$ times.

To understand the performance of the codes, we numerically evaluate the variances for random samples. We have considered random samples of three types---``realistic'', ``asymptotic'' and ``mixed''---to incorporate the generic properties of noises. The realistic samples take all $\log_{10}\left(\sigma_s\right)$ uniform in range $[-2, -0.7]$ to represent cases that are practically relevant; the asymptotic samples take all $\log_{10}\left(\sigma_s\right)$ uniform in range $[-3,-2]$ to probe the asymptotic performances; while the mixed samples take $\log_{10}\left(\sigma_s\right)$ uniform in each of the three different ranges: $[-4,-3]$, $[-3,-2]$ and $[-2,-1]$, to represent the case where noises are very different across the different channels.

First, we focus on the ultimate performance, by obtaining the minimum of the global-optimized noise, $\min_{\bm \Pi}\sigma_{L}^\star(\bm \Pi)$, and the minimum of the greedy-optimized noise $\min_{\bm \Pi}\sigma_{L}^\circledast(\bm \Pi)$. In Fig.~\ref{fig:sigma_scaling}(c)(d), we plot the global minimum noise (blue and red open circles) vs the geometric mean $\bar{\sigma}$ for random samples of noises with three and four modes, similar to the two-mode case. In all samples, we find a good agreement with the asymptotic result in Eq.~\eqref{eq:asymptotic_u} (black solid lines), which shows a power law $\bar{\sigma}^n$ (black dashed lines). The greedy scheme (blue and red open triangles) gives substantially worse performance as expected, which manifests the importance of choosing the proper encoding and decoding. For a benchmark, we also plot the ultimate STD lower bound in Ineq.~\eqref{sigma_L_limit} (gray stars and gray solid lines), which shows the same power-law with the geometric mean $\bar{\sigma}$. The gray star shows the precise lower bound for each sample, while the gray solid lines show the further lower bound of $\sigma_L\ge \bar{\sigma}^n/\sqrt{e}$ in Ineq.~\eqref{sigma_L_limit}.

\begin{table}[t]
\begin{tabular}{|l|l|l|l|l|l|}
\hline
Code    &noise  STDs    & $G_1$ & $G_2$ & $G_3$ & $\sigma_L^\star$ \\ \hline
\multirow{3}{*}{GKP-TMS}&\{0.1,0.1\}         &4.807&     &     &0.03580\\ 
&\{0.1,0.1,0.1\}     &3.541&6.949&     &0.01632\\ 
&\{0.1,0.1,0.1,0.1\} &3.037&5.376&7.041&0.008319\\ \hline
\multirow{3}{*}{GKP-SR}&\{0.1,0.1\}         &2.933&     &     &0.03583\\ 
&\{0.1,0.1,0.1\}     &2.532&2.752&     &0.01559\\ 
&\{0.1,0.1,0.1,0.1\} &2.242&2.529&2.533&0.007538\\ \hline
\end{tabular}
\caption{Data of the optimal gain $\bm G$ and the resulting logical noise STD $\sigma_L^\star$ for both GKP-TMS and GKP-SR codes. We choose a homogeneous noise model with an STD $\sigma=0.1$ for two, three and four modes.
\label{table:data}
}
\end{table}

Now we further look into the global minimized noise $\sigma_{L}^\star(\bm \Pi)$ as a function of the order $\bm \Pi$. 
Taking the random samples as examples, we calculate their relative ratios of $\sigma_{L}^\star(\bm \Pi)/\min_{\bm \Pi} \sigma_{L}^\star(\bm \Pi)$ for each permutation $\bm \Pi$. To represent the general case, we take the sample average for each permutation and plot the average relative ratios in Fig.~\ref{fig:permutations_3_4}. Although fluctuations exist on some of the permutations (see Fig.~\ref{fig:PermutationList_TMS} and Fig.~\ref{fig:PermutationList_SqzRep}), an average value close or equal to unity means that the corresponding permutation is optimal. For the 4-mode cases in Fig.~\ref{fig:permutations_3_4} (d), the relative ratios are generally greater when the most noisy channel is put on the bottom, such as orders 1, 3, 7 and 13. While relative ratios are smaller when the least noisy channel is put on the bottom, such as orders 12, 18 and 24. We see that in most cases the reverse order $\bm \Pi=(n,n-1,\cdots,2,1)$ gives a relative ratio of unity, therefore is the optimal order of permutation in most cases being considered, for both GKP-TMS and GKP-SR codes. After the global optimization of the encoding parameter $\bm G$, the overall differences between the permutations are small (less than one-order-of-magnitude), considering that the noise can be suppressed to many orders of magnitude smaller than the original noise. Indeed, suppose one optimizes over the entire Gaussian encoding unitary $\hat{U}_{\bm E, \bm 0}$ in Fig.~\ref{fig:general_code}, the permutations of the noisy channels will not affect the overall performance, as Gaussian unitary enables the arbitrary swapping between modes. Below, we give some intuition about the different permutations.

For the two-mode case of the GKP-TMS code, the numerical results show that the order $\bm \Pi=(1,2)$ works better than the other order  $\bm \Pi=(2,1)$. This is caused by effects from the side peaks beyond the main Gaussian peak in the noise distributions. As to multi-mode concatenation, because $\sigma_u \gg \sigma^{(u)}$ is true after the first layer, we can ignore the effects of side peaks in the noise distribution of higher-order concatenations i.e., $\sigma^{(u+1)}_L\approx\sigma^{(u+1)}$, then $\sigma^{(u+1)}_L\approx\sigma^{(u+1)}\approx {\sigma^{(u)}}/{\sqrt{G-1}}$  as shown in Appendix~\ref{App:TMS_concatenation}. So we prefer to put the less noisy channel on the bottom. Intuitively, the order $\bm \Pi=(n,n-1,\cdots,1,2)$ should work better.
Indeed, although the reverse order has good performances, the orders $(3,1,2)$ and $(4,3,1,2)$ are in fact as good as, if not slightly better than, the reverse orders $(3,2,1)$ and $(4,3,2,1)$ respectively. Namely, at the bottom layer, one can utilize the order $(1,2)$ identical to the two-mode case; while for the higher-order layers, the non-Gaussian distribution complicates the situation and the reversed orders turn out to have better performances.

For the two-mode case of the GKP-SR code, the numerical results show that the order $\bm \Pi=(2,1)$ works better. Similarly, for higher orders, if we ignore the effects of side peaks, we have $\sigma^{(u+1)}_L\approx\sigma^{(u+1)}\approx {\sigma^{(u)}}/{G}$ as shown in Appendix~\ref{App:sqz_rep}. So the order $\bm \Pi=(n,n-1,\cdots,2,1)$ is expected to work better intuitively. Since our asymptotic analysis does not include the effect of side peaks after the first layer, when the differences between $\sigma$'s are small, other orders may work as good or even slightly better sometimes.

\begin{figure}[t]
\centering
\includegraphics[width=0.45\textwidth]{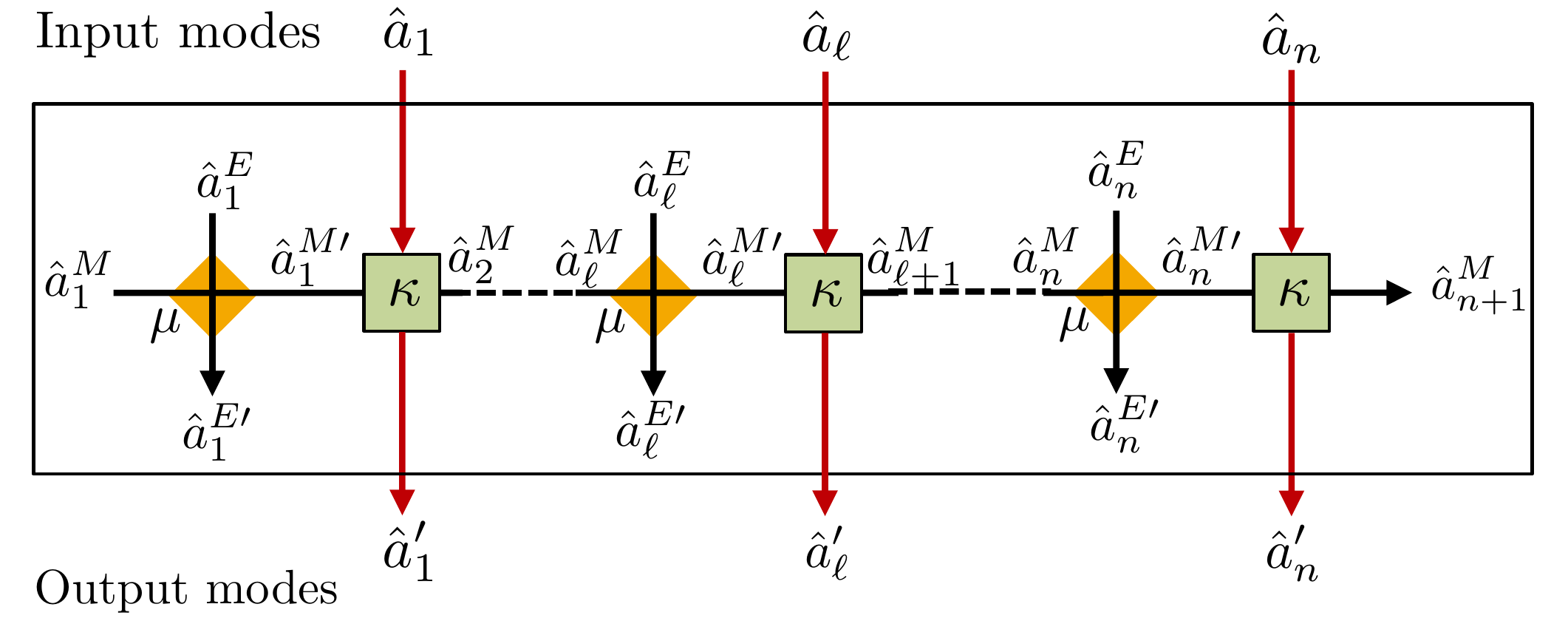}
\caption{Schematic of an n-mode loss channel with memory. The memory mode $\hat{a}_1^M$ is initially in vacuum. In the $\ell$-th channel use, the memory mode $\hat{a}_\ell^M$ mixes with a new environment mode $\hat{a}_\ell^E$ (in vacuum), on a beamsplitter with transmissivity $\mu$, to produce the environment mode $\hat{a}_\ell^{M\prime}$ for the $\ell$-th channel. The input mode $\hat{a}_\ell$ is mixed with $\hat{a}_\ell^{M\prime}$ on a beamsplitter with transmissivity $\kappa$, leading to the $\ell$-th output mode $\hat{a}_\ell^\prime$.
\label{schematic_memory_channel}
}
\end{figure}

Before closing, we look into the optimal parameter $\bm G$. For simplicity, we focus on the homogeneous case where $\sigma_\ell=\sigma$ for all $1\le \ell \le n$, so that the performance does not depend on the order of permutation. In Table~\ref{table:data}, we provide the optimal gains $\bm G$ for two, three and four modes, with the value $\sigma=0.1$ chosen. We see that as the number of layers increases, the gain at the initial layers decreases---the squeezing is completed by multiple steps, each with a smaller strength, such that the additional side-peaks created during the iteration are suppressed. Overall, despite the noise model being homogeneous, the optimal values of the gain are heterogeneous.

\begin{table}[t]
\centering
\begin{tabular}{|l|l|l|l|l|l|l|}
\hline
6 Modes   & $\sigma_1$ & $\sigma_2$ & $\sigma_3$ & $\sigma_4$ & $\sigma_5$ & $\sigma_6$ \\ \hline
STD       & 0.0792      & 0.0881    & 0.107     &  0.150    &  0.269    &   0.839  \\ 
\hline
\hline
Index & $\bm{\Pi}$  & $G_1$  & $G_2$  & $G_3$  & $G_4$  & $\sigma_L^\star$ \\ \hline
85    & (4,3,1,2,5) & 1.008 & 4.379 & 5.647 & 3.727 & 0.008652       \\ \hline
87    & (4,3,2,1,5) & 1.008 & 4.456 & 5.599 & 3.734 & 0.008681       \\ \hline
\end{tabular}
\caption{Top: table of the noise STDs in the memory channel. We have sorted the noises from small to large. Bottom: the best two permutations that minimize the noise STD.
\label{table:noise_memory}
}
\end{table}
\begin{figure}[t]
    \centering
    \includegraphics[width=0.45\textwidth]{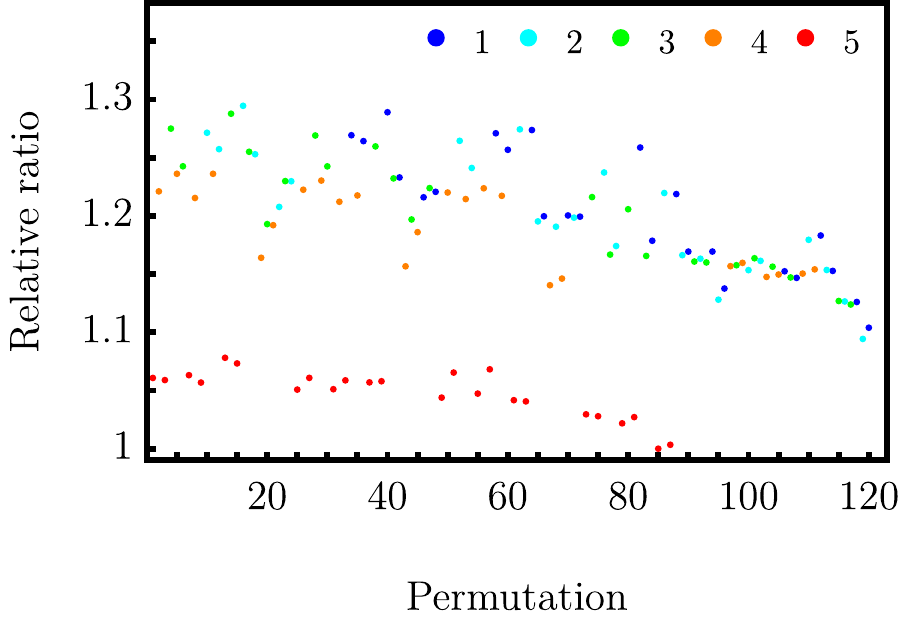}
    \caption{Relative ratio $\sigma_{L}^\star(\bm \Pi)/\min_{\bm \Pi} \sigma_{L}^\star(\bm \Pi)$ versus permutation $\bm{\Pi}$ for GKP-TMS code with five modes. The color indicates the mode on the bottom layer, for example, red color means the permutations with $5$ as the last integer, which puts the most noisy mode on the bottom layer. }
    \label{fig:relative-ratio-5}
\end{figure}

\section{Gaussian memory channels as an example}
\label{sec:memory_channel}

In the following, we consider a Gaussian memory channel as an example for the entire QEC design process. In a communication link, as the rate of data transmission increases, memory effects~\cite{lupo2010capacities,caruso2014quantum} will come into play, where the noises between different modes are correlated from the cross-talk between the environment modes.

As shown in Fig.~\ref{schematic_memory_channel}, a memory channel can be described by a sequence of thermal-loss channels with environmental modes correlated through a single memory mode $\hat{a}_1^M$. In fact, at each layer indexed by $1\le \ell \le n$, the memory mode $\hat{a}_\ell^M$ mixes with an environment mode $\hat{a}_\ell^E$, on a beamsplitter with transmissivity $\mu$, to produce the environment mode $\hat{a}_\ell^{M\prime}$ for the $\ell$-th channel. The input mode $\hat{a}_\ell$ is mixed with $\hat{a}_\ell^{M\prime}$ on a beamsplitter with transmissivity $\kappa$, leading to the $\ell$-th output mode $\hat{a}_\ell^\prime$. The overall channel takes input modes $\{\hat{a}_\ell\}_{\ell=1}^n$ and produces the output modes $\{\hat{a}_\ell^\prime\}_{\ell=1}^n$ in a correlated fashion. Ref.~\cite{lupo2010capacities} devices an $n$-mode passive linear optics transform before and after the $n$-mode memory channel to map the input and output to a suitable set of bases, such that the overall channel becomes an independent collection of pure loss channels $\{\calL_{\tau_\ell,0},1\le \ell \le n\}$ with
\be 
\tau_\ell=\abs{\frac{\sqrt{\mu}-\sqrt{\kappa}e^{i \pi \ell/n}}{1-\sqrt{\kappa \mu}e^{i \pi \ell/n}}}^2.
\ee 
An amplifier $\calA_{1/\tau_\ell}$ prior to each channel can reduce the overall channel to the form of Eq.~\eqref{noise_model}, with $\sigma_\ell = \sqrt{1-\tau_\ell}$. This procedure of channel reduction is precisely an example of Lemma~\ref{lemma:general_reduction}.

We apply our QEC design scheme for the $n = 6$ case of a memory channel with transmissivities $\mu = 0.9$ and $\kappa = 0.8$. The STDs of the channel noises after unraveling according to Lemma~\ref{lemma:general_reduction} are listed in Table~\ref{table:noise_memory}. To begin with, the $\Phi_{\sigma_6^2}$ channel is obviously too noisy to be used for error-correction, as it is well beyond the thresholds identified in Fig.~\ref{fig:contours}. As a consequence, we apply the concatenated GKP-TMS code to the remaining five channels. By optimizing the parameter $\bm{G}$ for each order $\bm{\Pi}$, we obtain the optimal STD $\sigma_L^*(\bm \Pi)$ of the corrected noise. Similar to Fig.~\ref{fig:permutations_3_4}, we plot the relative ratio with respect to the optimal STD to compare the performance of different orders in Fig.~\ref{fig:relative-ratio-5}. To gain further insights, we color each point according to which mode is being utilized at the bottom layer. As we can see, the red dots have the minimum noise levels, indicating that it's optimal to put the most noisy channel with the STD $\sigma_5= 0.2692$ at the bottom layer, which is consistent with the intuition. The best two orders are given by indices $85$ and $87$, which correspond to $\bm{\Pi} = (4,3,1,2,5)$ and $\bm{\Pi} = (4,3,2,1,5)$ respectively. As shown in Table~\ref{table:noise_memory} bottom panel, the squeezing factor $G_1$ is close to $1$ showing that the noisy channel $\Phi_{\sigma_5^2}$ at the bottom layer is almost discarded to achieve the best performance, as expected.

\section{Applications}
In this section, we consider the applications in Fig.~\ref{schematic_concept} of the CV QEC developed in this paper. As the application in distributed sensing has been considered in Ref.~\cite{zhuang2020distributed}, we will focus on the cases of network communication and quantum memory with heterogeneous storage efficiencies. In both scenarios, one wants to protect a quantum state with high fidelity, either in transmission or storage. Therefore, we calculate the fidelity improvement from QEC.

\begin{figure}[t]
    \centering
    \includegraphics[width=0.45\textwidth]{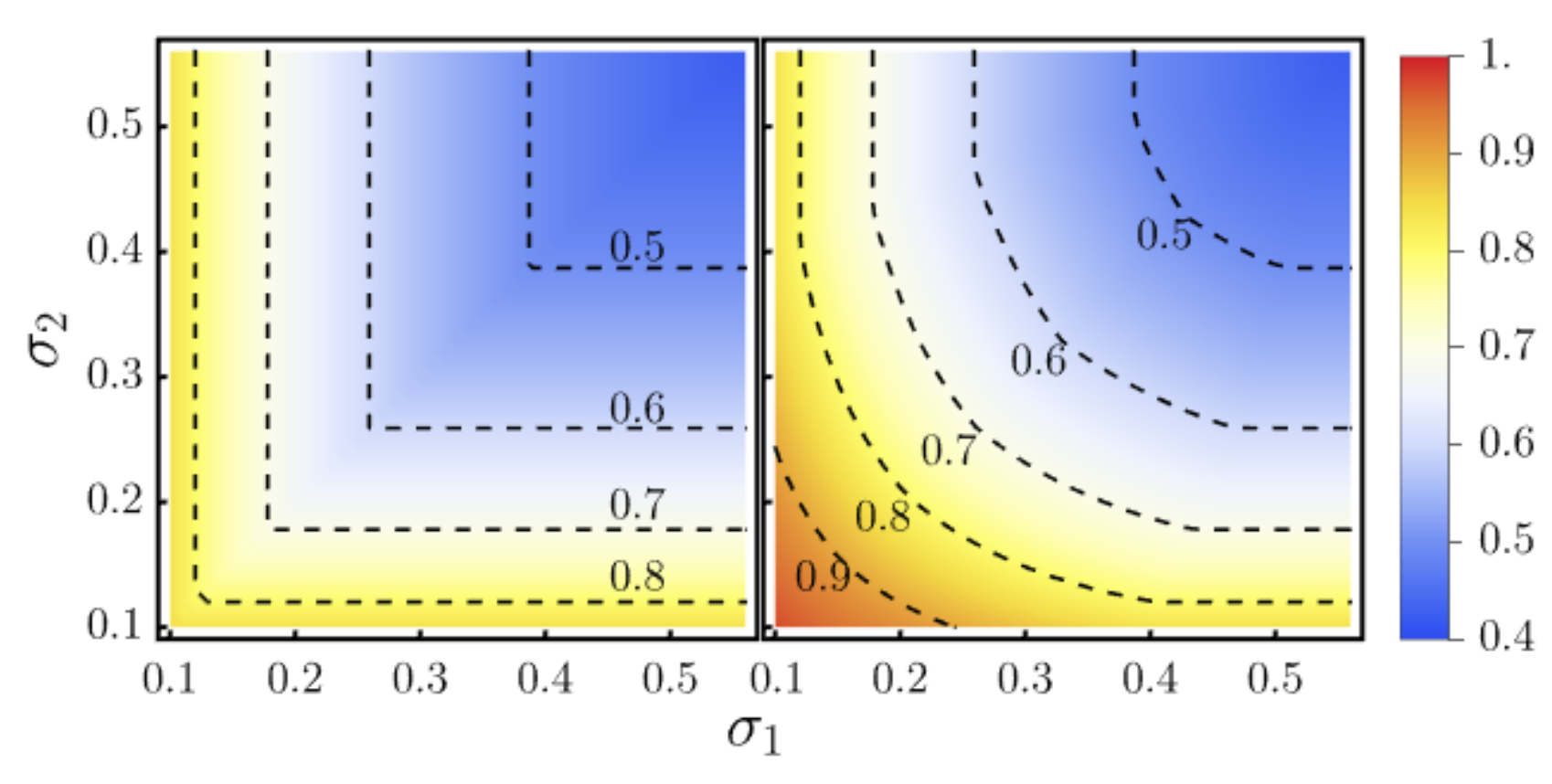}
    \caption{Contours of quantum fidelity transmitting a squeezed-vacuum state with $20$dB of squeezing (a) without error correction (b) with GKP-TMS code. }
    \label{fig:Fidelity-contour}
\end{figure}

To give an example, we consider the data mode to be a single-mode squeezed-vacuum state and apply the GKP-TMS code. We can calculate the fidelity when the data state goes through a AWGN channel analytically, and that after error correction numerically (see Appendix ~\ref{App:Fidelity}). Similar to Fig.~\ref{fig:contours}, we will pass the data state through the less noisy channel with STD $\min(\sigma_1,\sigma_2)$ and the QEC will take the optimal permutation.
Fig.~\ref{fig:Fidelity-contour} shows the quantum fidelities with and without QEC when transmitting a single-mode squeezed-vacuum state with $20$dB squeezing. We can see improvement in the fidelity in a wide range of the noise levels. And higher-order codes will be able to further improve it, similar to what we have shown for the noise variance in previous sections.

Before closing, we also address the experimental realization side of CV QEC design in this paper. 
In superconducting
microwave cavity systems, interactions are relatively easy to engineer. Indeed, GKP state engineering and application in QEC for qubits have been recently demonstrated~\cite{campagne2020quantum}. The system therein can be adapted to demonstrate the CV QEC proposed in this paper. In the optical domain, it is much more challenging as interactions between optical modes are much harder to engineer, due to the the requirement of inline two-mode squeezing and generation of the non-Gaussian GKP state. For the first difficulty, one can resort to methods of reducing inline squeezing to linear optics and offline squeezing, such as the one introduced in Ref.~\cite{filip2005}.

\section{Conclusions and discussions}
\label{sec:conclusion}

In this paper, we provide a QEC design strategy for correcting correlated Gaussian noises in bosonic systems. Our channel reduction scheme is able to map a general noise model to a collection of additive noise channels. The optimized concatenation of the two-mode codes achieves the optimal scaling of the logical noise vs the input noise.

Although the evaluation of the noise distribution is efficient, the exact calculation still requires precisely keeping track of all the Gaussian peaks in the noise distribution. It is natural to ask if an estimation can be obtained without doing so. One approach is to approximate the noise distribution output from each layer of QEC as Gaussian, so that each layer the error correction has the same setup of the first layer. In terms of the noise variance, this leads to the simple concatenation in Eq.~\eqref{eq:asymptotic_u}, which works well when the initial noises are small as confirmed in Fig.~\ref{fig:sigma_scaling} by the black solid lines. In terms of the fidelity, we calculate the fidelity assuming a Gaussian distribution in Appendix~\ref{App:Fidelity}, which works well in most values of noises. These results also indicate that a quick estimation is possible by assuming a Gaussian distribution of the output noise after the QEC.

Finally, we point out two future directions. First, the performance of the general GKP-Gaussian code introduced in Section~\ref{sec:general_QEC} is not solved. Second, it is an open problem how to generalize the noise model to incorporate common non-Gaussian noises such as phase noise.

\begin{acknowledgments}
This research project is supported by the Defense Advanced Research Projects Agency (DARPA) under Young Faculty Award (YFA) Grant No. N660012014029. 
\end{acknowledgments}


%

\appendix

\section{Proof of Lemma~\ref{lemma:general_reduction}: reduction of a general Gaussian channel}
\label{app:proof_reduction}

\subsection{Step 1: reduce a general multi-mode Gaussian channel to a correlated AWGN channel}

Denote a general Gaussian unitary $\hat{U}_{\bm S, \bm d}$ and the corresponding unitary channel as $\calU_{\bm S, \bm d}$, which corresponds to the Bogoliubov transform when acting on the quadrature operators $\hat{\bm x}=\left(\hat{q}_1,\hat{p}_1,\cdots,\hat{q}_n,\hat{p}_n\right)$
\be 
\hat{U}_{\bm S, \bm d}^\dagger \hat{\bm x} \hat{U}_{\bm S, \bm d}=\bm S \hat{\bm x}+\bm d.
\ee 

In terms of the mean $\bar{\bm x}$ and covariance matrix $\bm V$ of a single mode state, a thermal-loss channel $\calL_{\eta,N_B}$ corresponds to the transform
\be 
\bar{\bm x}\to \sqrt{\eta} \bar{\bm x}, \bm V \to \eta \bm V+(2N_B+1-\eta)\bm I_2,
\ee 
while a noisy amplifier channel $\calA_{G,N_B}$ corresponds to the transform
\be 
\bar{\bm x}\to \sqrt{G} \bar{\bm x}, \bm V \to G \bm V+(2N_B+G-1)\bm I_2.
\ee 
For later use, we also write out the explicit transform when the channel only acts on a single mode (e.g. the first one) among $n\ge 2$ modes, then we can write out the transforms in block forms
\begin{widetext}
\be 
\bar{\bm x}\to (\sqrt{\eta}\bar{\bm x}_{1,2},\bar{\bm x}_{3:2n}), \bm V\to
\left( \begin{array}{cc}
\sqrt{\eta}\bm I_2 & \bm 0  \\
\bm 0 & \bm I_{2n-2}  
\end{array} \right)
\bm V
\left( \begin{array}{cc}
\sqrt{\eta}\bm I_2 & \bm 0  \\
\bm 0 & \bm I_{2n-2}  
\end{array} \right)
+
\left( \begin{array}{cc}
(2N_B+1-\eta)\bm I_2 & \bm 0  \\
\bm 0 & \bm 0  
\end{array} \right),
\ee 
\be 
\bar{\bm x}\to (\sqrt{G}\bar{\bm x}_{1,2},\bar{\bm x}_{3:2n}), \bm V\to
\left( \begin{array}{cc}
\sqrt{G}\bm I_2 & \bm 0  \\
\bm 0 & \bm I_{2n-2}  
\end{array} \right)
\bm V
\left( \begin{array}{cc}
\sqrt{G}\bm I_2 & \bm 0  \\
\bm 0 & \bm I_{2n-2}  
\end{array} \right)
+
\left( \begin{array}{cc}
(2N_B+G-1)\bm I_2 & \bm 0  \\
\bm 0 & \bm 0  
\end{array} \right).
\ee
\end{widetext}
For the pure loss case where $N_B=0$, we also write the channels as $\calL_\eta$ and $\calA_G$ directly. 

A general $n$-mode Gaussian channel $\Phi_{\bm T, \bm N, \bm d}$~\cite{Weedbrook_2012,Holevo2001} can be described by a displacement vector $\bm d$ and two $2n\times 2n$ real matrices $\bm T, \bm N$, which satisfy
\be 
\bm N+i\bm \Omega-i\bm T \bm \Omega \bm T^T\ge 0,
\ee 
where $\bm \Omega=\bigoplus_{k=1}^n i\bm Y$ with $\bm Y$ being the Pauli matrix.
The action of the channel on a Gaussian state with mean $\bar{\bm x}$ and covariance matrix $\bm V$ can be described by
\be 
\bar{\bm x}\to \bm T\bar{\bm x}+\bm d, \bm V\to \bm T \bm V \bm T^T+\bm N.
\ee 
As $\bm T$ is a real matrix, we have the singular value decomposition $\bm T=\bm A^T \bm G^\prime \bm B^T$, with $\bm G^\prime$ as a $2n\times 2n$ diagonal matrix and $\bm A^T, \bm B^T$ as $2n\times 2n$ orthogonal matrices. As real orthogonal matrices corresponds to general passive linear transforms, we can design the following procedure. We pre-process the input state by a passive linear transform described by $\bm B$ and post-process the output state by a displacement $-\bm d$ and another passive linear transform described by $\bm A$. The overall transform  $\calU_{\bm A,-\bm d} \circ \Phi_{\bm T, \bm N, \bm d}\circ \calU_{\bm B,0} 
$ leads to the mean and covariance
\begin{align}
\bar{\bm x}\to&\bm A\left(\bm T \left( \bm B\bar{\bm x}\right)+\bm d-\bm d\right)
\nonumber
\\
&
=  \bm A\bm A^T \bm G^\prime \bm B^T \left( \bm B\bar{\bm x}\right)=\bm G^\prime \bar{\bm x},
\\
\bm V\to &\bm A \left(\bm T \bm B\bm V\bm B^T \bm T^T+\bm N\right) \bm A^T
\nonumber
\\
&=  \bm A \left(\bm A^T \bm G^\prime \bm B^T \bm B\bm V\bm B^T \bm B \bm G^\prime \bm A+\bm N\right) \bm A^T
\nonumber
\\
&
=\bm G^\prime \bm V \bm G^\prime +\bm A \bm N \bm A^T. 
\end{align}
Now we consider the transform described by $\bm G^\prime=\bigoplus_{k=1}^{2n} G_k^\prime$. We can apply a set of squeezing operations described by the symplectic transform
\be 
\bm S=\bigoplus_{k=1}^{n} {\rm Diag}\left(\sqrt{G_{2k}^\prime/G_{2k-1}^\prime},\sqrt{G_{2k-1}^\prime/G_{2k}^\prime}\right)
\ee 
in the post-processing, which leads to the overall transform $ \calU_{\bm S,0} \circ \calU_{\bm A,-\bm d} \circ \Phi_{\bm T, \bm N, \bm d}\circ \calU_{\bm B,0} 
$
described by
\begin{align}
&\bar{\bm x}\to\bm G \bar{\bm x}
\\
&\bm V\to \bm G \bm V \bm G +\bm S\bm A \bm N \bm A^T \bm S^T,
\end{align}
where now $\bm G=\bigoplus_{k=1}^{n} G_k \bm I_2$ with $G_k=\sqrt{G_{2k}^\prime G_{2k-1}^\prime}$.

For $G_k>1$, we post-process with a pure-loss channel $\calL_{1/G_k^2}$ on the $k$-th mode; while for $G_k<1$, we pre-process with a quantum-limited amplifier $\calA_{1/G_k^2}$ on the $k$-th mode. The pre-processing amplification is
\be 
\calA^{\rm pre}=\otimes_{k=1}^n \calA_{\max[1,1/G_k^2]}= \otimes_{k=1}^n \calA_{1/\min(G_k^2,1)},
\ee 
and the post-processing pure-loss is
\be 
\calL^{\rm post}=\otimes_{k=1}^n \calL_{\min[1,1/G_k^2]}=\otimes_{k=1}^n \calL_{1/\max(G_k^2,1)}.
\ee

Then the overall channel
$ 
\calL^{\rm post}\circ \calU_{\bm S,0} \circ \calU_{\bm A,-\bm d} \circ \Phi_{\bm T, \bm N, \bm d}\circ \calU_{\bm B,0} \circ \calA^{\rm pre}
$
leads to the overall mapping
\begin{widetext}
\begin{align}
&\bar{\bm x} \to \bar{\bm x},
\\
&
\bm V \to \bm V
+\left(\bigoplus_{k=1}^n \frac{1}{\max(G_k,1)}\bm I_2\right)\bm S\bm A \bm n \bm A^T \bm S^T \left(\bigoplus_{k=1}^n \frac{1}{\max(G_k,1)}\bm I_2\right)
+\bigoplus_{k=1}^n \max(0,1-G_k^2)\bm I_2
+\bigoplus_{k=1}^n \max(0,1-\frac{1}{G_k^2})\bm I_2.
\end{align}
\end{widetext}
Now we see that the overall channel has become a multi-mode AWGN channel with a potentially complicated covaraince matrix.

\subsection{Step 2: reduce a correlated AWGN channel to an independent collection of AWGN channels}

In general, the multi-mode AWGN channel acting on the input state $\hat{\rho}_{\hat{a}_1\cdots \hat{a}_n}$ leads to the output
\be 
\Sigma_{\bm W}\left(\hat{\rho}_{\hat{a}_1\cdots \hat{a}_n}\right)
=
\int d^{2n}\bm x  F_{\bm W}\left(\bm x\right) \hat{D}\left(\bm x\right)\hat{\rho}_{\hat{a}_1\cdots \hat{a}_n} \hat{D}^\dagger\left(\bm x\right),
\ee 
where the multi-mode extension follows the tensor rule: the operator $\hat{D}\left(\bm x\right)=\otimes_{\ell=1}^n \hat{D}\left(x_{2\ell-1},x_{2\ell}\right)$ and 
$F_{\bm W}\left(\bm x\right)$ is a Gaussian PDF with a $2n$-by-$2n$ covariance matrix $\bm W$.

One can apply two $n$-mode passive linear optics unitary channels $\calU_{\bm O,0}$ and $\calU_{\bm O,0}^\dagger = \calU_{\bm O^T,0}$, before and after the $n$-mode correlated channel to obtain the concatenation
\be 
\calU_{\bm O^T,0}\circ \Sigma_{\bm W} \circ \calU_{\bm O,0}=\Sigma_{\bm O^T\bm W \bm O},
\ee 
such that the new covariance matrix $\bm O^T\bm W \bm O={\rm Diag}\left(W_1,\cdots, W_{2n}\right)$ is diagonal. Note that because the original AWGN is symmetric between the position and momentum quadratures, the new covariance matrix is also symmetric. Therefore, Lemma~\ref{lemma:general_reduction} is proven.

\section{General lower bounds on the logical noise variances}
\label{proof_lower_bound}

From Theorem S1 of Ref.~\cite{noh2019encoding}, the quantum capacity $C_Q(\calN_{Q,P})$ of an additive non-Gaussian noise channel $\calN_{Q,P}$ is lower bounded by
\be 
C_Q(\calN_{Q,P})\ge \max[0,\log_2\left(\frac{1}{e\sigma_q \sigma_p}\right)],
\label{LB}
\ee 
where $\sigma_p$ and $\sigma_q$ are the STDs of the momentum and position quadratures.
Then Ref.~\cite{noh2019encoding} derived the ultimate STD lower bound for the error correction of GKP-two-mode-squeezing code. Here we generalize to the case of multiple heterogeneous noise case.

The overall quantum communication utilizes the channel $\otimes_{\ell=1}^n \Phi_{N_\ell=\sigma_\ell^2}$, with variances $N_\ell=\sigma_\ell^2$ on both quadratures. The quantum capacity of a single-mode additive thermal noise channel is upper bounded~\cite{noh2018quantum,rosati2018narrow,sharma2018bounding} by
\be 
C_Q\left(\Phi_{\sigma^2}\right)\le \log_2 \left(\frac{1-\sigma^2}{\sigma^2}\right).
\ee 
We can use the quantum capacity of a pure loss channel to easily generalize this upper bound. Note that when any $\sigma_\ell>1$, we will discard the channel immediately, as the channel has zero quantum capacity; when $\sigma_\ell=0$, then we immediately choose that channel for communication and there is no need for error correction; therefore, we only need to consider the case when $0< \sigma_\ell <1$.

We know that a pure-loss channel is anti-degradable ($\eta<1/2$) and weak-degradable ($\eta>1/2$)~\cite{caruso2006degradability}, therefore utilizing the proof technique in Ref.~\cite{devetak2005capacity}, the quantum capacity is additive between different pure-loss channels
\be 
C_Q(\otimes_\ell \calL_{\eta_\ell})= \sum_\ell C_Q(\calL_{\eta_\ell})=\sum_\ell \log_2\left(\frac{\eta_\ell}{1-\eta_\ell}\right),
\ee 
where in the last step we utilized the (unconstrained) capacity formula resulting from Refs.~\cite{Holevo2001,wolf2007quantum,noh2018quantum,caruso2006degradability}. Now we utilize the channel decomposition $\Phi_{1-\eta}=\calL_\eta \circ \calA_{1/\eta}$ and $\Phi_{\frac{1-\eta}{\eta}}= \calA_{1/\eta}\circ \calL_\eta$, which leads to the upper bound
\begin{align}
C_Q\left(\Phi_{\sigma^2}\right)&\le \min \left[C_Q\left(\calL_{1-\sigma^2}\right), C_Q\left(\calL_{\frac{1}{1+\sigma^2}}\right)\right]
\nonumber
\\
&=C_Q\left(\calL_{1-\sigma^2}\right).
\end{align}
Similarly, we can have
\begin{align}
C_Q\left(\otimes_{\ell=1}^n\Phi_{\sigma^2_\ell}\right)&\le C_Q\left(\otimes_{\ell=1}^n \calL_{1-\sigma^2_\ell}\right)
\nonumber
\\
&=\sum_{\ell=1}^n\log_2 \left(\frac{1-\sigma^2_\ell}{\sigma^2_\ell}\right).
\label{UB}
\end{align}

In the error correction procedure, all channels $\otimes_{\ell=1}^n\Phi_{\sigma^2_\ell}$ are utilized to transmit a single mode with a reduced logical noise, therefore combining Ineqs.~\eqref{LB} and~\eqref{UB}, we have
\be
\log_2\left(\frac{1}{e\sigma_q \sigma_p}\right)\le \sum_{\ell=1}^n\log_2 \left(\frac{1-\sigma^2_\ell}{\sigma^2_\ell}\right).
\ee 
A simple loose bound can be obtained by further upper bounding $\log_2 \left(\frac{1-\sigma^2_\ell}{\sigma^2_\ell}\right)< \log_2 \left(\frac{1}{\sigma^2_\ell}\right)$ and therefore
\be 
\sigma_q\sigma_p\ge \frac{1}{e} \prod_{\ell=1}^n \frac{\sigma^2_\ell}{1-\sigma^2_\ell}> \frac{1}{e} \prod_{\ell=1}^n \sigma_\ell^2.
\ee 
The above inequality gives the ultimate performance of the error correction scheme. We can also introduce
\be
\sigma_L^2\equiv \frac{1}{2}(\sigma_q^2+\sigma_p^2)\ge \sigma_p\sigma_q\ge \frac{1}{e} \prod_{\ell=1}^n \frac{\sigma^2_\ell}{1-\sigma^2_\ell}\ge  \frac{1}{e} \prod_{\ell=1}^n \sigma_\ell^2.
\ee

\section{Details of GKP-TMS code with heterogeneous AWGN noises}

\label{App:TMS_code}

The joint PDF of $\boldsymbol{z}  = (z_q^{(1)},z_p^{(1)},z_q^{(2)},z_p^{(2)})$ is given by
\begin{align}
    \label{eq:A1}
    \begin{split}
     &
     P(z_q^{(1)},z_p^{(1)},z_q^{(2)},z_p^{(2)}) = \frac{1}{(2\pi)^{2} |\boldsymbol{V}_{\boldsymbol{z}}|^{\frac{1}{2}}}\exp \left( -\frac{1}{2} \boldsymbol{z}^{T}\boldsymbol{V}_{\boldsymbol{z}}^{-1}\boldsymbol{z}\right)\\
     &
     =\frac{1}{(2\pi)^2\sigma_1^2\sigma_2^2}\exp\left\{-\frac{1}{2}\left\{
     \left[(G-1)\sigma_1^2+G\sigma_2^2\right]\left(z_q^{(1)2}+z_q^{(2)2}\right)
    \right.
    \right.\\
     &
     +\left[G\sigma_1^2+(G-1)\sigma_2^2\right]\left(z_p^{(1)2}+z_p^{(2)2}\right)+\\
     &
     \left.
     \left.
     2\sqrt{G(G-1)}(\sigma_1^2+\sigma_2^2)(z_q^{(1)}z_q^{(2)}-z_p^{(1)}z_p^{(2)})
     \right\}
     \right\}.
    \end{split}
\end{align}
Then the conditional distribution is:
\begin{widetext}
\begin{align}
    \begin{split}
        & P(z_q^{(1)},z_p^{(1)}|z_q^{(2)},z_p^{(2)}) =
        \frac{P(z_q^{(1)},z_p^{(1)},z_q^{(2)},z_p^{(2)})}{P(z_q^{(2)},z_p^{(2)})} = \frac{P(z_q^{(1)},z_p^{(1)},z_q^{(2)},z_p^{(2)})}{\int^{\infty}_{-\infty} P(z_q^{(1)},z_p^{(1)},z_q^{(2)},z_p^{(2)})\diff{z_q^{(1)}}\diff{z_p^{(1)}}}
        \\
        & = \frac{(G-1)\sigma_1^2+G\sigma_2^2}{2\pi\sigma_1^2\sigma_2^2}
        \exp\left\{ -\frac{1}{2\sigma_1^2\sigma_2^2}\left\{
        \left[\sqrt{\left(G-1\right)\sigma_1^2+G\sigma_2^2}z_q^{(1)}+\sqrt{\frac{G(G-1)}{\left(G-1\right)\sigma_1^2+G\sigma_2^2}}(\sigma_1^2 + \sigma_2^2)z_q^{(2)}\right]^2+
        \right. \right.
        \\
        & 
        \left.
        \left.
        \left[
            \sqrt{(G-1)\sigma_1^2+G\sigma_2^2}
            z_p^{(1)}
            -\sqrt{\frac{G(G-1)}{(G-1)\sigma_1^2+G\sigma_2^2}}
            \left(\sigma_1^2 + \sigma_2^2\right)
            z_p^{(2)}
        \right]^2
        \right\}
        \right\}.
    \end{split}
\end{align}
\end{widetext}

Here, we have two types of estimators ($\bar{z}_q^{(1)}$, $\bar{z}_p^{(1)}$) based on the mesurement result ($z_q^{(2)}$, $z_p^{(2)}$). The first is the maximum likelihood estimator (MLE)
\begin{align}
    \label{eq:4}
    (\bar{z}_q^{(1)}, \bar{z}_p^{(1)}) = \arg_{(z_q^{(1)}, z_p^{(1)})} \: \max\: P({z}_q^{(1)},{z}_p^{(1)}|{z}_q^{(2)},{z}_p^{(2)}).
\end{align}
The second approach is to minimize the variance, which leads to using the average as the estimator as
\begin{align}
    \label{eq:5}
    (\bar{z}_q^{(1)}, \bar{z}_p^{(1)}) = \braket{({z}_q^{(1)},{z}_p^{(1)})}_{ P(\cdot,\cdot|{z}_q^{(2)},{z}_p^{(2)})},
\end{align}

Therefore, from Eq.~\eqref{eq:5} we get
\begin{align}
    \begin{split}
    &
    \bar{z}_q^{(1)} = \braket{z_q^{(1)}}_{P(\cdot,\cdot|z_q^{(2)},z_p^{(2)})} =   -\frac{\sqrt{G(G-1)}(\sigma_1^2+\sigma_2^2)}{(G-1)\sigma_1^2+G\sigma_2^2}z_q^{(2)},
    \\
    &
     \bar{z}_p^{(1)} = \braket{z_p^{(1)}}_{P(\cdot,\cdot|z_q^{(2)},z_p^{(2)})} =   \frac{\sqrt{G(G-1)}(\sigma_1^2+\sigma_2^2)}{(G-1)\sigma_1^2+G\sigma_2^2}z_p^{(2)}.
     \end{split}
\end{align}
Because the PDF $P(z_q^{(1)},z_p^{(1)}|z_q^{(2)},z_p^{(2)})$ is Gaussian, it achieves its maximum at its mean value. Therefore, the two estimations in Eq.~\eqref{eq:4} and Eq.~\eqref{eq:5} agree with each other. Now the probability density functions of logical quadrature noises $\xi_q$ and $\xi_p$ can be derived from the joint PDF. Since the measurements on $\ket{\rm{GKP}}$ state have the module $\sqrt{2\pi}$ uncertainty, the corrected quadrature noises are given by
\begin{align}
    \begin{split}
        &
        \xi_q = z_q^{(1)} - \bar{z}_q^{(1)} 
        \nonumber
        \\
        &= z_q^{(1)} + \frac{\sqrt{G(G-1)}(\sigma_1^2+\sigma_2^2)}{(G-1)\sigma_1^2+G\sigma_2^2}R_{\sqrt{2\pi}}(z_q^{(2)}),
        \\
        &
        \xi_p = z_p^{(1)} - \bar{z}_p^{(1)} 
        \nonumber
        \\
        &= z_p^{(1)} - \frac{\sqrt{G(G-1)}(\sigma_1^2+\sigma_2^2)}{(G-1)\sigma_1^2+G\sigma_2^2}R_{\sqrt{2\pi}}(z_p^{(2)}).
    \end{split}
\end{align}
Similar to Eq.~\eqref{eq:A1}, from the covariance matrix in Eq.~\eqref{covMatrix2},
we can get the joint PDF of $z_q^{(1)}$ and $z_q^{(2)}$ as
\begin{align}
    &
    P(z_q^{(1)}, z_q^{(2)}) =
    \nonumber
    \\
    &\frac{1}{2\pi\sigma_1\sigma_2}\exp\left\{-\frac{1}{2\sigma_1^2\sigma_2^2}\left\{[(G-1)\sigma_1^2+G\sigma_2^2]{z_q^{(2)}}^2 + \right.\right.
    \nonumber
    \\
    &
    \quad
    +[G\sigma_1^2+(G-1)\sigma_2^2]{z_q^{(2)}}^2 
    \nonumber
    \\
    &\left.\left.
    + 2\sqrt{G(G-1)}(\sigma_1^2 + \sigma_2^2)z_q^{(1)}z_q^{(2)}\right\}\right\}.
\end{align}
Then the PDF of $\xi_q$ is given by:
\begin{align}
    \begin{split}
        Q(\xi_q) &= \int_{-\infty}^\infty\diff{z_q^{(1)}}\int_{-\infty}^\infty\diff{z_q^{(2)}}P(z_q^{(1)},z_q^{(2)})
        \nonumber
        \\
        &\times\delta\left[\xi_q-z_q^{(1)}-\frac{\sqrt{G(G-1)}(\sigma_1^2+\sigma_2^2)}{(G-1)\sigma_1^2+G\sigma_2^2}R_{\sqrt{2\pi}}(z_q^{(2)})\right]\\
        &
        = \sum_{n\in\mathbb{Z}}b_n \; F_{\frac{\sigma_1\sigma_2}{\sqrt{(G-1)\sigma_1^2+G\sigma_2^2}}}(\xi_q+\mu_n),
    \end{split}
\end{align}
where the coefficients and the means are
\begin{align}
    \begin{split}
        &b_n = \int_{(n-\frac{1}{2})\sqrt{2\pi}}^{(n+\frac{1}{2})\sqrt{2\pi}}\diff{z} F_{\sqrt{(G-1)\sigma_1^2+G\sigma_2^2}}(z),\\
        &
        \mu_n = \frac{\sqrt{G(G-1)}(\sigma_1^2+\sigma_2^2)}{(G-1)\sigma_1^2+G\sigma_2^2}\sqrt{2\pi}n,
    \end{split}
\end{align}
and
$F_\sigma$ is the PDF of a zero-mean Gaussian distribution with a variance $\sigma^2$.
Following the same steps, we can also show that the residue noise $\xi_p$ obey the same statistics as $\xi_q$. The output variance is therefore
\begin{align}
    \label{eq:nthorderSTD}
    \sigma_L^2 = \frac{\sigma_1^2\sigma_2^2}{(G-1)\sigma_1^2+G\sigma_2^2} + \sum_{n \in \mathbb{Z}}b_n\mu_n^2.
\end{align}
We also derive the asymptotic expressions of the optimal gain that minimizes the output variance when both $\sigma_1$ and $\sigma_2$ are small.
The leading order of the above noise variance can be obtained as
\begin{align}
    &\sigma_L^2 \simeq \frac{\sigma_1^2\sigma_2^2}{(G-1)\sigma_1^2+G\sigma_2^2} 
    \nonumber
    \\
    &+ \frac{2\pi G(G-1)(\sigma_1^2+\sigma_2^2)^2}{{[(G-1)\sigma_1^2+G\sigma_2^2]}^2} {\rm Erfc}\left(\frac{\sqrt{\pi}}{2\sqrt{(G-1)\sigma_1^2+G\sigma_2^2}}\right).
    \label{eq:1stroderSTD}
\end{align}
For a fixed gain G and $\sigma_2<\sigma_1$, Eq.~\eqref{eq:1stroderSTD} leads to
\begin{align}
    \sigma_L^2 \simeq  \frac{\sigma_1^2\sigma_2^2}{\sigma_G^2} + \frac{2\pi G(G-1)(\sigma_1^2+\sigma_2^2)}{\sigma_G^4} {\rm Erfc}\left(\frac{\sqrt{\pi}}{2\sigma_G}\right).
\end{align}
The above equation decreases as $\sigma_G$ increases as long as $\sigma_G\gtrsim0.7$, when $\sigma_1^2+\sigma_2^2$ and $G$ are fixed. In the small noise limit, we have $\sigma_G\gg 1$ and this condition is satisfied. Therefore, when $\sigma_2<\sigma_1\ll1$, we have
\begin{align}
    &
    \sigma_G(\sigma_1,\sigma_2)^2=(G-1)\sigma_1^2+G\sigma_2^2 
    \nonumber
    \\
    &
    < (G-1)\sigma_2^2+G\sigma_1^2 = \sigma_G(\sigma_2,\sigma_1)^2.
    \label{AppendixC11}
\end{align}
So we can switch the two channels to get a smaller STD of the output noise, then adjust the gain $G$ for a even smaller STD of the output noise after switching.

Numerical results show that the optimal $G\gg1$ when both $\sigma_1$ and $\sigma_2$ are small. Thus, $G \approx G-1$ and ${G(G-1)(\sigma_1^2+\sigma_2^2)^2} \approx {[(G-1)\sigma_1^2+G\sigma_2^2]}^2$.
Also the argument inside the Erfc function is much larger than 1. With $ x \equiv 1/{\sigma_G^2}$, Eq.~\eqref{eq:1stroderSTD} can be further simplified to
\begin{align}
    \begin{split}
        &
        \sigma_L^2= f(x)  \equiv \sigma_1^2\sigma_2^2x+2\pi\;{\rm Erfc}(\frac{\sqrt{\pi x}}{2}).
    \end{split}
\end{align}
The optimum $x^*$ can be found by solving
\be 
f'(x^*) = \sigma_1^2\sigma_2^2 +2\pi\left[-\frac{1}{2\sqrt{x^*}}\exp\left(-\frac{\pi x^*}{4}\right)\right] = 0,
\ee 
which leads to
\be 
x^* =  \frac{4}{\pi}\ln\left(\frac{\pi}{\sigma_1^2\sigma_2^2\sqrt{x^*}}\right).
\label{eq:A9}
\ee 
We solve Eq.~\eqref{eq:A9} by plugging in $x^*$ iteratively,
\begin{align}
    \begin{split}
        &
        x^* = \frac{4}{\pi}\ln\left(\frac{\pi}{\sigma_1^2\sigma_2^2\sqrt{x}}\right)\\
        &
        \quad
        = \frac{4}{\pi}\ln\left(\frac{\pi}{\sigma_1^2\sigma_2^2}\right)-\frac{2}{\pi}\ln(x^*)\\
        &
        \quad
        = \frac{4}{\pi}\ln\left(\frac{\pi}{\sigma_1^2\sigma_2^2}\right)-\frac{2}{\pi}\ln\left[\frac{4}{\pi}\ln\left(\frac{\pi}{\sigma_1^2\sigma_2^2}\right)-\frac{2}{\pi}\ln(x^*)\right]\\
        &
        \quad
        \approx \frac{4}{\pi}\ln\left(\frac{\pi}{\sigma_1^2\sigma_2^2}\right)-\frac{2}{\pi}\ln\left(\frac{4}{\pi}\right)-\frac{2}{\pi}\ln\left[\ln\left(\frac{\pi}{\sigma_1^2\sigma_2^2}\right)\right]\\
        &
        \quad
        \approx \frac{4}{\pi}\ln\left(\frac{\pi^{3/2}}{2\sigma_1^2\sigma_2^2}\right).
    \end{split}
\end{align}
Then the optimal gain and the output noise are given by
\begin{align}
    \begin{split}
        &
        G^*=\frac{1/x^*+\sigma_1^2}{\sigma_1^2+\sigma_2^2}=\frac{\frac{\pi}{4}\left[\ln\left(\frac{\pi^{3/2}}{2\sigma_1^2\sigma_2^2}\right)\right]^{-1}+\sigma_1^2}{\sigma_1^2+\sigma_2^2},
        \\
        &
        {\sigma_L^*}^2 \simeq \sigma_1^2\sigma_2^2x^*+2\pi \;{\rm Erfc}\left(\frac{\sqrt{\pi x}}{2}\right)\\
        &
        \quad
        \approx \sigma_1^2\sigma_2^2x^*+\frac{4}{\sqrt{x^*}} \exp\left(-\frac{\pi x^*}{4}\right)\\
        &
        \quad
        \approx
        \frac{4\sigma_1^2\sigma_2^2}{\pi}\left\{\ln\left(\frac{\pi^{3/2}}{2\sigma_1^2\sigma_2^2}\right)+\left[\ln\left(\frac{\pi^{3/2}}{2\sigma_1^2\sigma_2^2}\right)\right]^{-\frac{1}{2}}\right\}\\
        &
        \quad
        \approx
        \frac{4\sigma_1^2\sigma_2^2}{\pi}\ln\left(\frac{\pi^{3/2}}{2\sigma_1^2\sigma_2^2}\right).
    \end{split}
\end{align}
As numerical results show that $\sigma_L(\sigma_1,\sigma_2)$ is smaller when $\sigma_1<\sigma_2$, we get the asymptotic curve for $\sigma_L = t \sigma_1$ ($t < 1$) under this approximation by
\begin{align}
    \label{eq:TMS-asymptotic-curves}
    t^2 = \frac{4\sigma_2^2}{\pi}\ln\left(\frac{\pi^{3/2}}{2\sigma_1^2\sigma_2^2}\right).
\end{align}

\section{Details of the concatenation of GKP-TMS code}
\label{App:TMS_concatenation} 

The derivation is done by induction. At the $(u+1)$-th layer, the noises are independent random variables $(\xi_q^{(1)},\xi_p^{(1)},\xi_q^{(2)},\xi_p^{(2)} )$, which corresponds to the new mode introduced and a second mode from the $u$-th layer. The new mode has a Gaussian noise PDF
\begin{align}
Q(\xi_q^{(1)}) = P(\xi_p^{(1)}) = F_{\sigma_u}(\xi^{(1)}_{q,p}).
\end{align}
Suppose the second mode from the $u$-th layer has a PDF as a sum of Gaussian functions 
\begin{align}
    Q(\xi_q^{(2)}) = P(\xi_p^{(2)}) = \sum_{k\in \mathbb{Z}} b_k F_{\sigma^{(u)}}(\xi_{q,p}^{(2)}+t_k)
    \label{eq:TMS-u-layer-assumption}
\end{align}
where $F_{\sigma}(\cdot)$ is the PDF of a zero-mean Gaussian distribution with STD $\sigma$ and $\sum_{k \in \mathbb{Z}} b_k = 1$ are normalized. We also assume that the coefficients are symmetric
\begin{align}
 b_k=b_{-k},\\
 t_k = -t_{-k}.
\end{align}
Following the same procedures of the two-mode squeezing code with heterogeneous independent noises, we have the joint PDF of q and p:
\begin{widetext}
\begin{equation}
\begin{split}
    &
    P_{12}(\vec{\xi}) = F_{\sigma_u}(\xi_q^{(1)})F_{\sigma_u}(\xi_p^{(1)})\sum_{k_1,k_2 \in \mathbb{Z}} b_{k_1} b_{k_2} F_{\sigma^{(u)}}(\xi_{q}^{(2)}+t_{k_1}) F_{\sigma^{(u)}}(\xi_{p}^{(2)}+t_{k_2}),
    \\
    &
    P_{12}^{\prime}(\vec{z}) = P_{12}(S \vec{\xi}) = F_{\sigma_u}(\sqrt{G}z_q^{(1)}+\sqrt{G-1}z_q^{(2)})F_{\sigma_u}(\sqrt{G}z_p^{(1)}-\sqrt{G-1}z_p^{(2)}) \times \\ 
    & \qquad \sum_{k_1,k_2 \in \mathbb{Z}} b_{k_1} b_{k_2} F_{\sigma^{(u)}}(\sqrt{G-1}z_q^{(1)}+\sqrt{G}z_q^{(2)}+t_{k_1}) F_{\sigma^{(u)}}(-\sqrt{G-1}z_p^{(1)}+\sqrt{G}z_p^{(2)}+t_{k_2}).
\end{split}
\end{equation}
By separating q and p and reformulating them, we are able to find the estimators of $z_q^{(1)}$ and $z_p^{(1)}$ in terms of the second mode. 
\begin{align}
    \label{eq:TMS-joint-distribution-q}
    &
    P_{12,q}^{\prime}(z_q^{(1)},z_q^{(2)}) = \sum_{k\in \mathbb{Z}} b_k F_{\sigma_u}(\sqrt{G}z_q^{(1)}+\sqrt{G-1}z_q^{(2)})F_{\sigma^{(u)}}(\sqrt{G-1}z_q^{(1)}+\sqrt{G}z_q^{(2)}+t_k),\\
    &
     P_{12,p}^{\prime}(z_p^{(1)},z_p^{(2)})=\sum_{k \in \mathbb{Z}} b_k F_{\sigma_u}(\sqrt{G}z_p^{(1)}-\sqrt{G-1}z_p^{(2)}) F_{\sigma^{(u)}}(-\sqrt{G-1}z_p^{(1)}+\sqrt{G}z_p^{(2)}+t_k).
\end{align}
Eq.~\eqref{eq:TMS-joint-distribution-q} can be reformulated to:
\begin{align}
\label{eq:TMS-reformulated-joint-distribution-q}
    P_{12,q}^{\prime}(z_q^{(1)},z_q^{(2)})&=\sum_{k\in \mathbb{Z}} b_k \frac{1}{2\pi \sigma_u \sigma^{(u)}}\exp\left\{ -\frac{G \sigma^{(u)^2}+(G-1)\sigma_u^2}{2\sigma_u^2 \sigma^{(u)^2}}\left[z_q^{(1)}+  \notag  \right. \right.\\
    &
    \left.\left.
    \frac{\sqrt{G(G-1)}(\sigma_u^2+\sigma^{(u)^2})}{(G-1)\sigma_u^2+G\sigma^{(u)^2}}\left(z_q^{(2)}+\frac{\sigma_u^2}{\sqrt{G}(\sigma_u^2+\sigma^{(u)^2})} t_k\right)\right]^2\right\} \exp{-\frac{(z_q^{(2)}+\sqrt{G}t_k)^2}{2\left[(G-1)\sigma_u^2+G\sigma^{(u)^2}\right]}}.
\end{align}
\end{widetext}
Now we address the choice of the estimator and the residue distribution the noise after error-correction. We choose the estimator by MLE on the main peak of Eq.~\eqref{eq:TMS-reformulated-joint-distribution-q},
\begin{equation}
    \begin{split}
        & \bar{z}_q^{(1)} = -A_q R_{\sqrt{2\pi}}\left(z_q^{(2)}\right),\\
        & \bar{z}_p^{(1)} = A_q R_{\sqrt{2\pi}}\left(z_p^{(2)}\right),
    \end{split}
\end{equation}
where we have introduced the following notations to make things more compact,
\begin{align}
    &
    A_q = \frac{\sqrt{G(G-1)}(\sigma_u^2+\sigma^{(u)^2})}{(G-1)\sigma_u^2+G\sigma^{(u)^2}},\\
    &
    \sigma^{(u+1)} = \frac{\sigma_u \sigma^{(u)}}{\sigma_3},\\
    &
    \sigma_3 = \sqrt{(G-1)\sigma_u^2+G\sigma^{(u)^2}}.
\end{align}
After the correction, the additive noise of q is a sum of two random variables  
\begin{align}
    &
    \xi_q^\prime = z_q^{(1)} - \bar{z}_q^{(1)} = z_q^{(1)} + A_q R_{\sqrt{2\pi }}\left(z_q^{(2)}\right), \\
    &
    Q(\xi_q^\prime) = \int_{-\infty}^\infty\diff{z_q^{(1)}}\int_{-\infty}^\infty\diff{z_q^{(2)}} P_{12,q}^\prime(z_q^{(1)},z_q^{(2)}) 
    \nonumber
    \\
    &\times\delta\left[\xi_q-z_q^{(1)}-A_q R_{\sqrt{2\pi}}(z_q^{(2)})\right].
\end{align}
Finally we have the PDF for $\xi_q^\prime$:
\begin{align}
    \label{eq:TMS-update-rule-q}
    Q(\xi_q^\prime) &= \sum_{k,\ell \in \mathbb{Z}}\left[b_k \int_{(\ell-\frac{1}{2})\sqrt{2\pi}}^{(\ell+\frac{1}{2})\sqrt{2\pi}}
     F_{\sigma_3}(z_3+\sqrt{G}t_k)\diff{z_3} \right]
     \nonumber
     \\
    &\times
    F_{\sigma^{(u+1)}}\left[ \xi_q^\prime+A_q(\sqrt{2\pi}\ell+\frac{\sigma_u^2}{\sqrt{G}(\sigma_u^2+\sigma^{(u)^2})}t_k)\right] \notag \\
    &\equiv \sum_{k,\ell \in \mathbb{Z}}b_{k,\ell}F_{\sigma^{(u+1)}}\left(\xi_q^\prime+t_{k,\ell}\right).
\end{align}
Similarly, the PDF for $\xi_p^\prime$ is given by:
\begin{align}
    \label{eq:TMS-update-rule-p}
    P(\xi_p^\prime) &= \sum_{k,\ell \in \mathbb{Z}}\left[b_k \int_{(\ell-\frac{1}{2})\sqrt{2\pi}}^{(\ell+\frac{1}{2})\sqrt{2\pi}}
    F_{\sigma_3}(z_3+\sqrt{G}t_k) \diff{z_3}\right] 
    \nonumber
     \\
    &\times
    F_{\sigma^{(u+1)}}\left[ -\xi_p^\prime+A_q(\sqrt{2\pi}\ell+\frac{\sigma_u^2}{\sqrt{G}(\sigma_u^2+\sigma^{(u)^2})}t_k)\right] \notag \\
    &\equiv \sum_{k,\ell \in \mathbb{Z}}b_{k,\ell}F_{\sigma^{(u+1)}}\left(\xi_p^\prime-t_{k,\ell}\right).
\end{align}
\begin{figure*}
    \centering
    \subfigure[]{
    \includegraphics[width=0.35\textwidth]{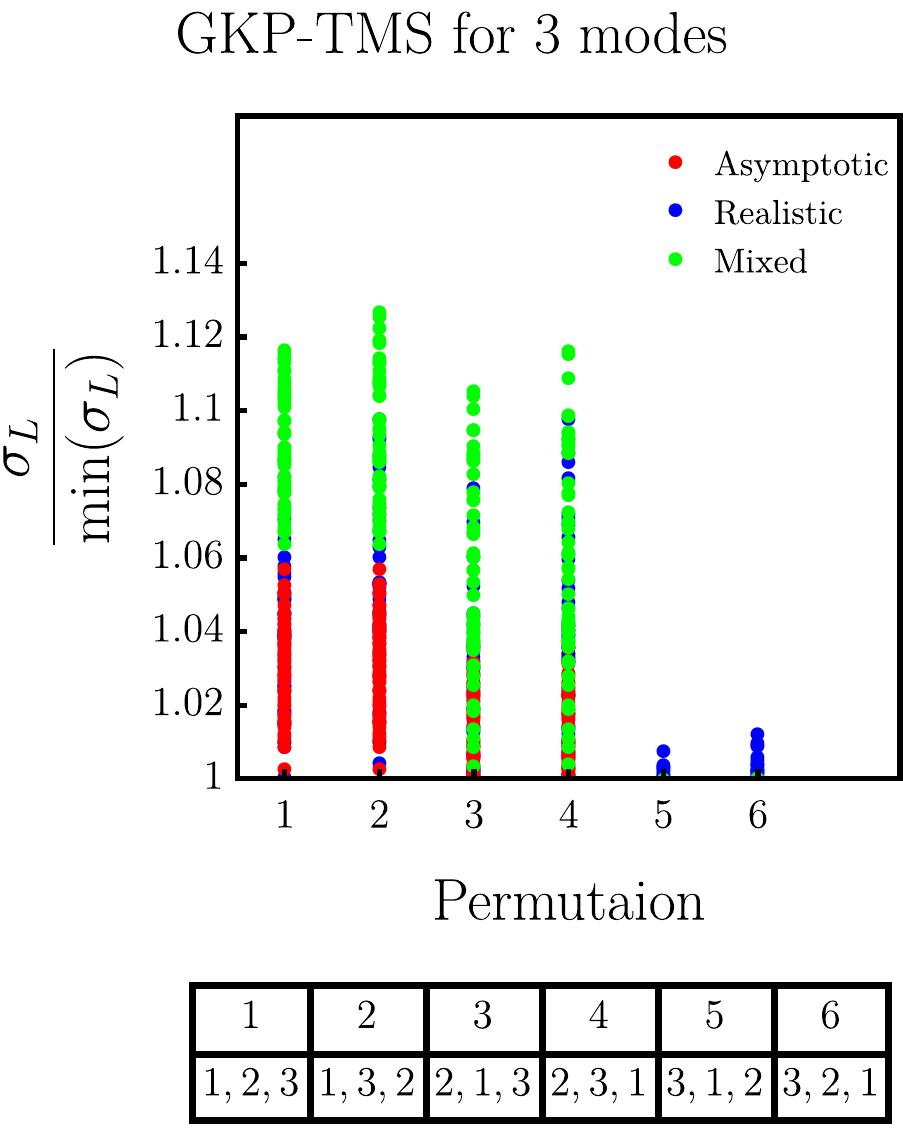}
    \label{fig:PermutationList_TMS3}
    }
    \subfigure[]{
    \includegraphics[width=0.35\textwidth]{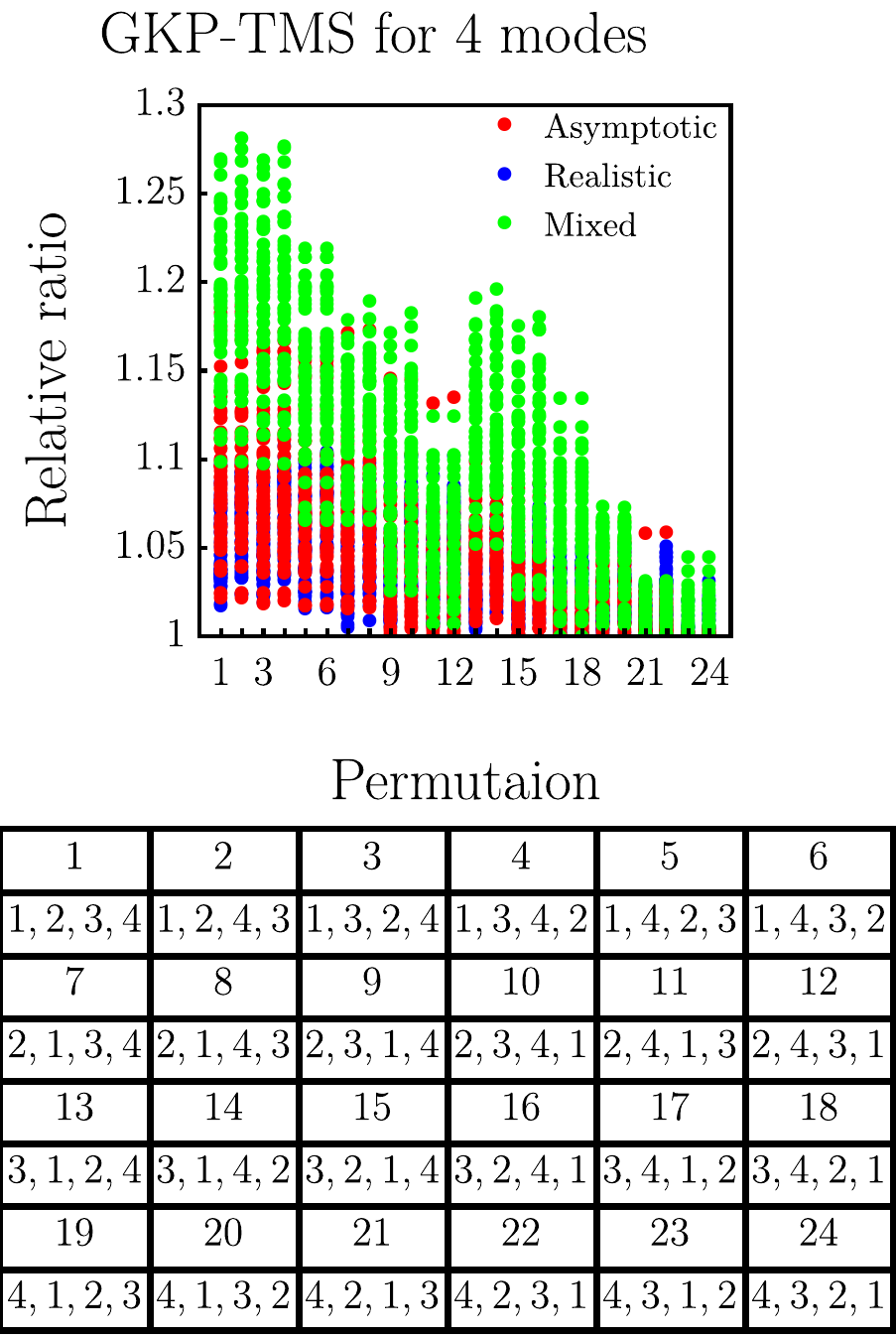}
    \label{fig:PermutationList_TMS4}
    }
    \caption{ Performance of the GKP-TMS code. The relative ratios of the STD for each permutation over the minimum STD for 3 (a) and 4 (b) modes. Each data point comes from the random samples, divided into three categories: asymptotic, realistic and mixed. The permutations are given in the tables on the bottom.
    \label{fig:PermutationList_TMS}
    }
\end{figure*}
From the assumption in Eq.~\eqref{eq:TMS-u-layer-assumption}, we see that $b_{k,\ell} = b_{-k,-\ell}$ and $t_{k,\ell}=-t_{-k,-\ell}$. Therefore the PDFs after the $(u+1)$-round are still symmetric. The $Q(\xi_q^\prime)$ and $P(\xi_p^\prime)$ have the same form of function as a sum of Gaussian functions. As we know at the bottom layer, the assumption of Eq.~\eqref{eq:TMS-u-layer-assumption} is true; therefore, by induction we have proven that in all layers, the PDFs satisfy the form in Eq.~\eqref{eq:TMS-u-layer-assumption}.

Let us perform asymptotic analyses by assuming $\sigma_u \gg \sigma^{(u)}$. This is usually a good approximation after the first layer of concatenation. Then $\sigma^{(u+1)}\approx {\sigma^{(u)}}/{\sqrt{G-1}}$. The leading term $\sigma^{(u+1)}$ decreases with $\sigma^{(u)}$. By induction, we should expect the order $(n,n-1,...,3,1,2)$ works well.
Note that in the first layer, $(1,2)$ is better than $(2,1)$. Next we show that the QEC code reduces the noise to a level of $\sigma^{(u+1)} \sim \prod_{k=0}^u \sigma_{u}$ asymptotically. On the one hand, we would like to take $\sigma_3 = c \sqrt{2\pi}$ for $c\ll 1$ so that the integral in Eq.~\eqref{eq:TMS-update-rule-q} takes most of its values around the origin. On the other hand, the smaller $\sigma_3$, the larger $\sigma^{(u+1)}$ is. Thus $\sigma^{(u+1)}={\sigma^{(u)}\sigma_u}/{(c \sqrt{2\pi})} \sim \sigma^{(u)}\sigma_u$ after optimizing $G$. By induction, it's expected that $\sigma^{(u+1)} \sim \prod_{k=0}^u \sigma_{u}$.

Fig.~\ref{fig:PermutationList_TMS3} and Fig.~\ref{fig:PermutationList_TMS4} show the numerical results of permutation plots for the three-mode case and the four-mode case respectively.

\section{Details of GKP-SR code}
\label{App:sqz_rep}

\begin{figure*}
\centering
\includegraphics[width=0.85\textwidth]{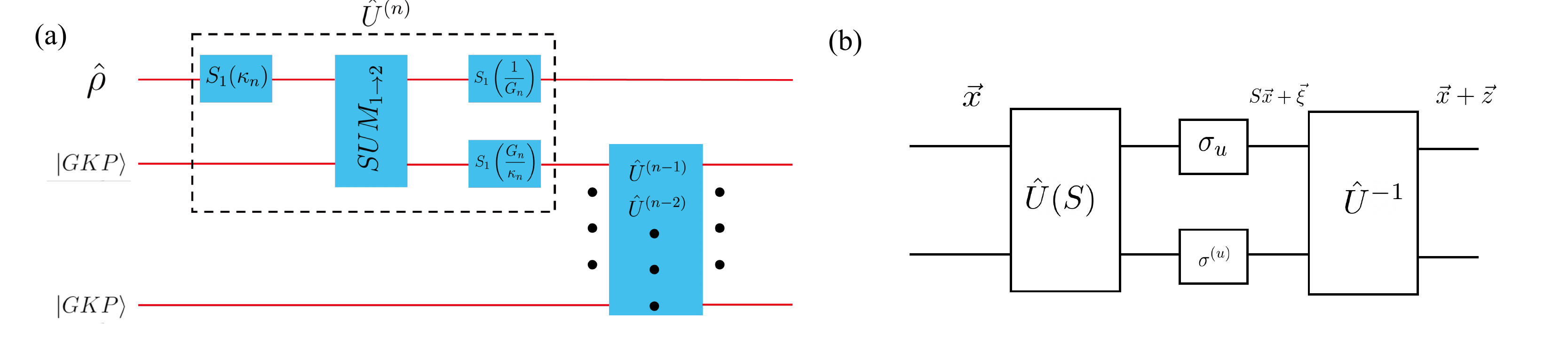}
\caption{
Schematic plots.
(a)Encoding of the GKP-SR code.
(b) The model of two channels.
\label{fig:encodingSqzRep}
}
\end{figure*}

This section solves the update rule of the noise characters at each layer for the GKP-SR code. In particular, we show that the noise after an arbitrary number of error-correction layers has a PDF composed of a sum of Gaussian functions. Let's first start with the symplectic transform of the GKP-SR code. As shown in Fig.~\ref{fig:encodingSqzRep}, the corresponding symplectic transform of the unitary $\hat{U}$ is
\begin{align}
    \label{eq:sqz-rep-transform-matrix}
    &
    S = \begin{pmatrix}
        \frac{\kappa}{G_{u+1}} && 0 && 0 && 0\\
        0 && \frac{G_{u+1}}{\kappa} && 0 && -G_{u+1}\\
        G_{u+1} && 0 && \frac{G_{u+1}}{\kappa} && 0\\
        0 && 0 && 0 && \frac{\kappa}{G_{u+1}}
        \end{pmatrix}.
\end{align}
The transform of the quadrature operator $\Vec{x} = (\hat{q}_1,\hat{p}_1,\hat{q}_2,\hat{p}_2)^T$ in the encoding, noisy transmission and decoding is as follows: $\Vec{x}\rightarrow S\Vec{x} \rightarrow S\Vec{x}+\Vec{\xi} \rightarrow \Vec{x}+S^{-1}\Vec{\xi}$ where $\Vec{\xi} = (\xi_q^{(1)},\xi_p^{(1)},\xi_q^{(2)},\xi_p^{(2)})$ is the independent AWGN noises. Therefore, at the output side before the measurement, the noise is transformed to $\Vec{z} = S^{-1}\Vec{\xi}$.

We prove the conclusion recursively. Consider the $(u+1)$-th layer of error-correction, where the noise input from the $u$-th layer has its PDF composed of a sum of Gaussian distributions, i.e.,
\begin{align}
&P^{(u)}_2\left(\xi_q^{(2)},\xi_p^{(2)}\right) = Q^{(u)}(\xi_q^{(2)})\times P^{(u)}(\xi_p^{(2)}),
\\
&
Q^{(u)}(\xi_q^{(2)})=\sum_{k\in \mathbb{Z}}b_{k,q}^{(u)}\;F_{\sigma^{(u)}}(\xi_q^{(2)}-t_{k,q}^{(u)}), \\
&
P^{(u)}(\xi_p^{(2)})=\sum_{k\in \mathbb{Z}}b_{k,p}^{(u)}\;F_{\sigma^{(u)}}(\xi_p^{(2)}-t_{k,p}^{(u)}),
\end{align}
where $F_\sigma(\cdot)$ is the zero-mean Gaussian PDF with STD $\sigma$. We have assumed the q and p quadratures can in general have different means $\{t_{k,q}^{(u)}\}$ and $\{t_{k,p}^{(u)}\}$. The superposition coefficients $b_{n,q}^{(u)}$ and $b_{n,p}^{(u)}$ of the quadratures can also be different. Here we will also assume the symmetry
\begin{align}
\label{eq:symmetry}
    b_{k,q}^{(u)} &=b_{-k,q}^{(u)},  &t_{k,q}^{(u)} &=-t_{-k,q}^{(u)},
    \\
    b_{k,p}^{(u)}&=b_{-k,p}^{(u)},   &t_{k,p}^{(u)}&=-t_{-k,p}^{(u)},
\end{align}
for $k =0,\pm1,\pm2,\cdots$. Also note that 
\begin{align}
    t_{0,q}^{(u)}=t_{0,p}^{(u)}=0.
\end{align}

In the $(u+1)$-th layer, we introduce an additional mode, which goes through an AWGN channel with variance $\sigma_u^2$. The PDF of the additional noise $\xi_q^{(1)},\xi_p^{(1)}$ is therefore
\begin{align}
    P_1\left(\xi_q^{(1)},\xi_p^{(1)}\right) = F_{\sigma_u}\left(\xi_q^{(1)}\right) \times F_{\sigma_u}\left(\xi_p^{(1)}\right).
\end{align}
This results in the overall noise PDF as a product
\be 
P_{12}\left(\vec{\xi}\right)=P_1\left(\xi_q^{(1)},\xi_p^{(1)}\right)\times P^{(u)}_2\left(\xi_q^{(2)},\xi_p^{(2)}\right).
\ee 
From the noise transformation $\Vec{z} = S^{-1}\Vec{\xi}$, before the measurement the joint noise PDF equals 
\begin{align*}
    P_{12}^\prime\left(\:\Vec{z}\:\right) 
    &= P_{12}\left(\:S\Vec{z}\:\right)\\
    &= F_{\sigma_u}\left(\frac{\kappa}{G_{u+1}}z_1\right) 
    F_{\sigma_u}\left(\frac{G_{u+1}}{\kappa}z_2-G_{u+1} z_4\right) \times
    \nonumber
    \\
    &
    \sum_{k_1\in \mathbb{Z}}b_{k_1,q}^{(u)}\;F_{\sigma^{(u)}}(G_{u+1} z_1 + \frac{G_{u+1}}{\kappa}z_3-t_{k_1,q}^{(u)})\times
    \nonumber
    \\
    &
    \sum_{k_2\in \mathbb{Z}}b_{k_2,p}^{(u)}\;F_{\sigma^{(u)}}(\frac{\kappa}{G_{u+1}}z_4-t_{k_2,p}^{(u)}).
\end{align*}

It can be seen from the above equation that at the output side, $z_1$ only correlates with $z_3$ and $z_2$ only correlates with $z_4$. We may write the joint PDFs of position and momentum separately, for example,
\begin{widetext}
\begin{align}
    P_{12, q}^\prime(z_1,z_3) 
    &=F_{\sigma_u}(\frac{\kappa}{G_{u+1}}z_1)
    \sum_{k=0,\pm1,\cdots}b_{k,q}^{(u)}\;F_{\sigma^{(u)}}(G_{u+1} z_1 + \frac{G_{u+1}}{\kappa}z_3-t_{k,q}^{(u)})
    \nonumber
    \\
    &=\frac{1}{2\pi\sigma_u\sigma^{(u)}} 
     b_{0,q}^{(u)}
    \exp{-\frac{1}{2}\left\{\frac{1}{\sigma_3^2}\left[z_1 + A_q z_3\right]^2+
    \frac{1}{\sigma_4^2}(z_3)^2 \right\}}
    \label{eq:main_peak}
    \\
    &+
    \frac{1}{2\pi\sigma_u\sigma^{(u)}} 
    \sum_{k=\pm1,\cdots} b_{k,q}^{(u)}
    \exp{-\frac{1}{2}\left\{\frac{1}{\sigma_3^2}\left[z_1 + A_q(z_3-\frac{\kappa}{G_{u+1}}t_{k,q}^{(u)})\right]^2+
    \frac{1}{\sigma_4^2}(z_3-\frac{\kappa}{G_{u+1}}t_{k,q}^{(u)})^2 \right\}}.
    \label{eq:sumpm1}
\end{align}
\end{widetext}
Here we have introduced the notations
\begin{align}
    &\sigma_3 \equiv \frac{G_{u+1}\sigma_u\sigma^{(u)}}{\sqrt{\kappa^2 \sigma^{(u)2}+G_{u+1}^4\sigma_u^2}},\\
    &
    \sigma_4 \equiv
    \frac{\sqrt{\kappa^2 \sigma^{(u)2}+G_{u+1}^4\sigma_u^2}}{G_{u+1}},\\
    &
    A_q \equiv \frac{G_{u+1}^4\sigma_u^2}{\kappa(\kappa^2\sigma^{(u)2}+G_{u+1}^4\sigma_u^2)}.
\end{align}
We have separated the PDF into two parts, the first part is the main peak in Eq.~\eqref{eq:main_peak}, and the second part is the rest of the peaks in Eq.~\eqref{eq:sumpm1}.

We will measure the GKP state of the second mode to infer about the noises $z_3,z_4$, from which we produce estimates of $z_1,z_2$ and the corresponding corrections on the noise of the first mode. Denote the measurement results $\tilde{z}_3, \tilde{z}_4$. Suppose we focus on the main peak in Eq.~\eqref{eq:main_peak}, the MLE and the minimum variance estimation both give
\begin{align}
    \bar{z}_1 = -A_q \tilde{z}_3.
    \label{estimator_z3}
\end{align}
It turns out that choosing this estimator will also simplify the update rule significantly.
Considering the ambiguity in the measurement result $\tilde{z}_3$ due to the GKP state, after applying a displacement $-\bar{z}_1$ to reduce the noise, we have the residual noise
\begin{align}
    \xi'_q = z_1-\bar{z}_1 
    = z_1+A_q R_{\sqrt{2 \pi}}(z_3).
\end{align}
The PDF of the position quadrature of the output in the $(u+1)$-th layer is therefore
\begin{widetext}
\begin{align*}
    Q^{(u+1)}(\xi'_q)    
    &= \int_{-\infty}^\infty\diff{z_1}\int_{-\infty}^\infty\diff{z_3}P_{12,q}^\prime(z_1,z_3) \times\delta\left[\xi'_q-z_1-A_q R_{\sqrt{2\pi}}(z_3)\right]\\
    &=\sum_{k,\ell \in \mathbb{Z}}b_{k,q}^{(u)} F_{\sigma_3}\left(\xi'_q+A_q(\ell\sqrt{2 \pi}-\frac{\kappa}{G_{u+1}}t_{k,q}^{(u)})\right) \times \int_{(\ell-\frac{1}{2})\sqrt{2\pi}}^{(\ell+\frac{1}{2})\sqrt{2\pi}}\diff{z_3}F_{\sigma_4}(z_3-\frac{\kappa}{G_{u+1}}t_{k,q}^{(u)}).
\end{align*}
\end{widetext}
Note that if one does not choose this estimator in Eq.~\eqref{estimator_z3}, then in the above equation, one will have additional cross terms between $\xi^\prime_q z_3$. This coupling leads to non-Gaussian function of $\xi^\prime_q$ in general.
The output momentum PDF can be calculated in a similar way using the mean value of the main peak as
\begin{widetext}
\begin{align*}
    P^{(u+1)}(\xi'_p) = \sum_{k,\ell \in \mathbb{Z}} b_{k,p}^{(u)}\; F_{\frac{\kappa\sigma_u}{G_{u+1}}}\left(\xi'_p-\kappa \ell \sqrt{2 \pi}\right)\times \int_{(\ell-\frac{1}{2})\sqrt{2\pi}}^{(\ell+\frac{1}{2})\sqrt{2\pi}}\diff{z_3}F_{\frac{G_{u+1}\sigma^{(u)}}{\kappa}}\left(z_3-\frac{G_{u+1}}{\kappa}t_{k,p}^{(u)}\right).
\end{align*}
\end{widetext}
We choose $\kappa$ to balance the variances of the main peaks of both quadratures:
\begin{figure*}
    \centering
    \subfigure[]{
    \includegraphics[width=0.35\textwidth]{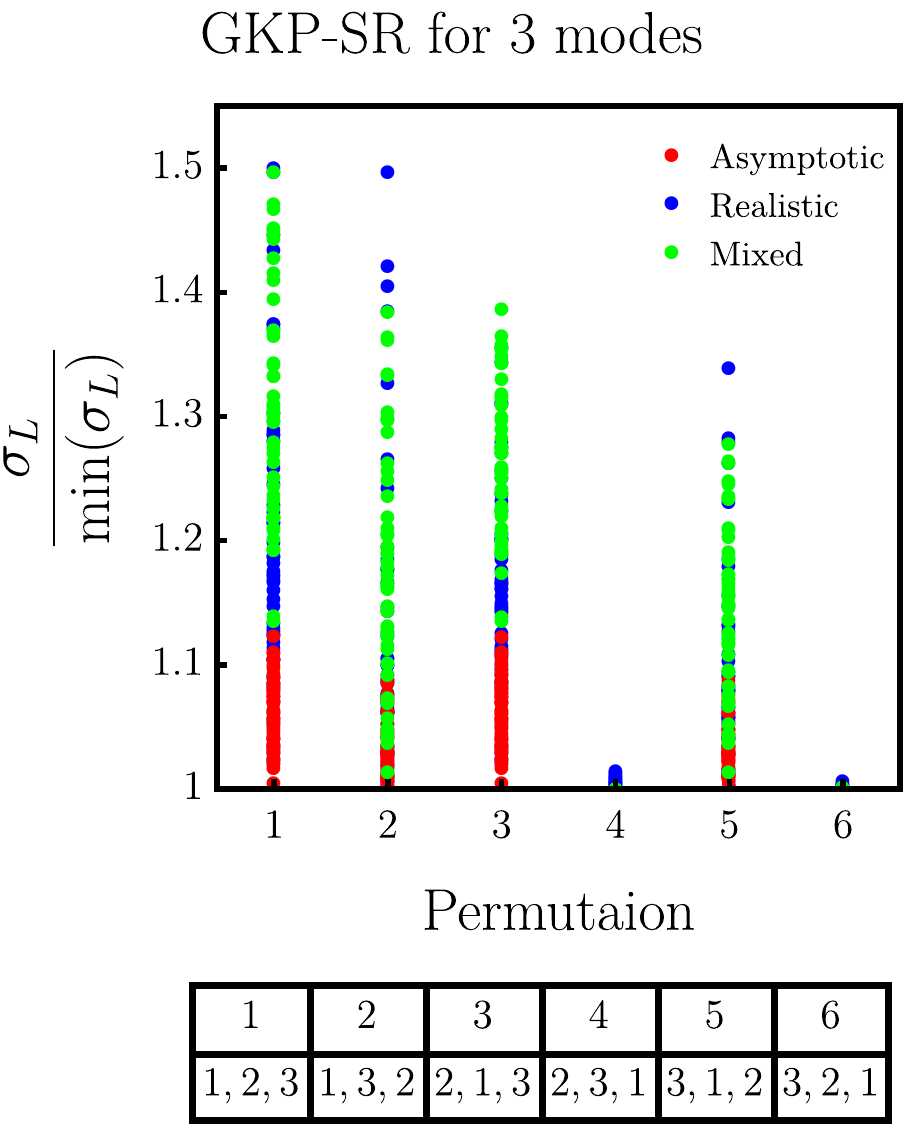}
     \label{fig:PermutationList_SqzRep3}
     }
    \subfigure[]{
    \includegraphics[width=0.35\textwidth]{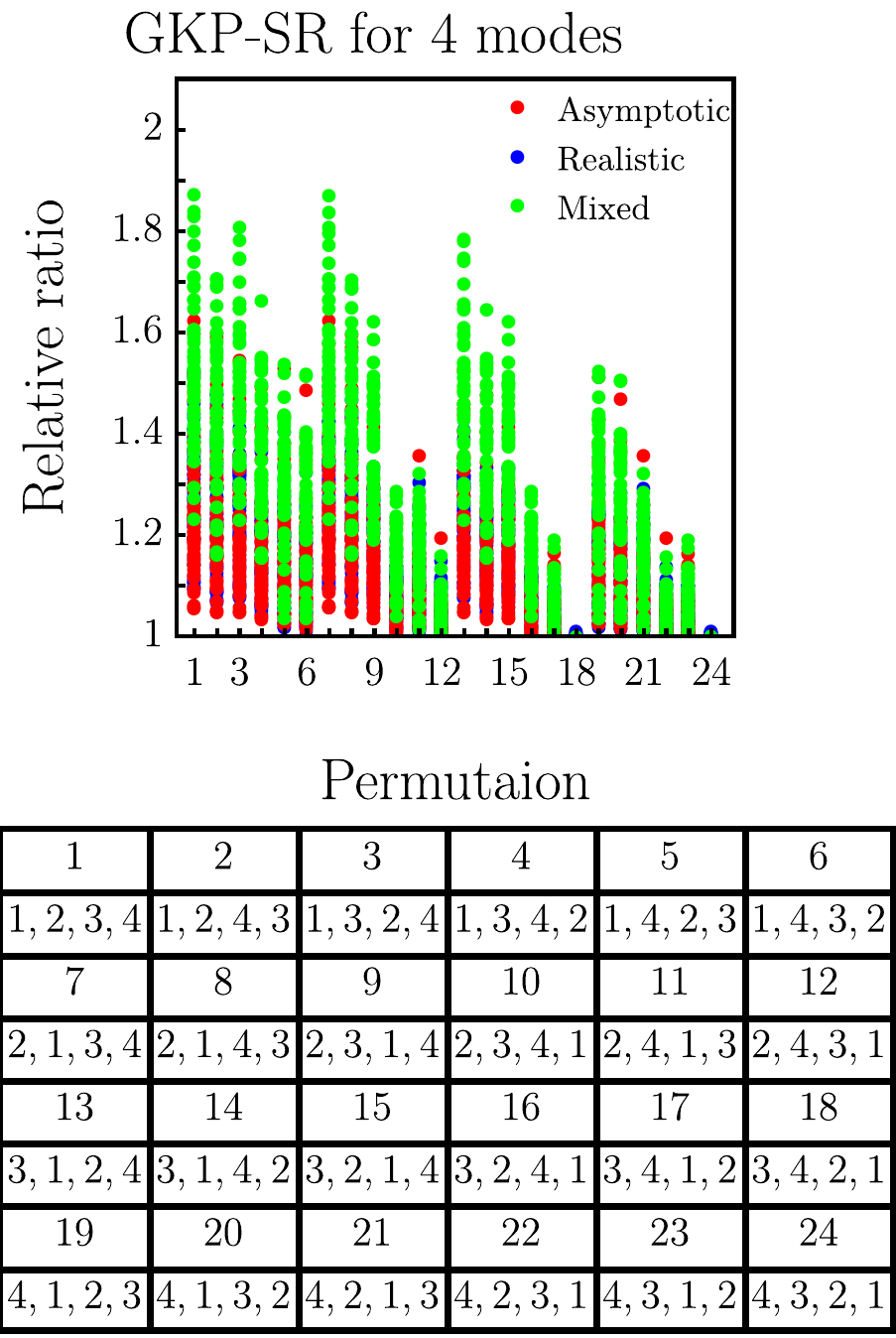}
     \label{fig:PermutationList_SqzRep4}
    }
    \caption{Performance of the GKP-SR code. The relative ratios of the STD for each permutation over the minimum STD for 3 (a) and 4 (b) modes. Each data point comes from the random samples, divided into three categories: asymptotic, realistic and mixed. The permutations are given in the tables on the bottom.
    \label{fig:PermutationList_SqzRep}
    }
\end{figure*}
\begin{widetext}
\begin{align}
    \label{eq:kappa}
    \frac{\kappa \sigma_u}{G_{u+1}} = \frac{G_{u+1} \sigma_u \sigma^{(u)}}{\sqrt{\kappa^2 \sigma^{(u)2} + G_{u+1}^4 \sigma_u^2}} 
    \Rightarrow
    \kappa^2 = \frac{\sqrt{G_{u+1}^8 \sigma_u^4+ 4 G_{u+1}^4\sigma^{(u)4}}-G_{u+1}^4\sigma_u^2}{2\sigma^{(u)2}}.
\end{align}
With this choice of $\kappa$, the PDFs are simplified to the following,
\begin{align}
    \label{eq:sq-rep PDFs1}
    \begin{split}
    Q^{(u+1)}(\xi'_q) &= \sum_{k,\ell \in \mathbb{Z}}b_{k,q}^{(u)}\; F_{\frac{\kappa \sigma_u}{G_{u+1}}}\left[\xi'_q+\kappa\frac{\sigma_u}{\sigma^{(u)}}(\ell\sqrt{2 \pi}-\frac{\kappa}{G_{u+1}}t_{k,q}^{(u)})\right] 
    \times
    \int_{(\ell-\frac{1}{2})\sqrt{2\pi}}^{(\ell+\frac{1}{2})\sqrt{2\pi}}\diff{z_3}F_{\frac{G_{u+1}\sigma^{(u)}}{\kappa}}(z_3-\frac{\kappa}{G_{u+1}}t_{k,q}^{(u)})\\
    &\equiv
    \sum_{k'\in \mathbb{Z}} b_{k',q}^{(u+1)} F_{\sigma_{u+1}}(\xi'_q-t_{k',q}^{(u+1)}),
    \end{split}
    \\
    \label{eq:sq-rep PDFs2}
    \begin{split}
        P^{(u+1)}(\xi'_p) &= \sum_{k,\ell \in \mathbb{Z}} b_{k,p}^{(u)}\; F_{\frac{\kappa\sigma_u}{G_{u+1}}}\left(\xi'_p-\kappa \ell \sqrt{2 \pi}\right)\times \int_{(\ell-\frac{1}{2})\sqrt{2\pi}}^{(\ell+\frac{1}{2})\sqrt{2\pi}}\diff{z_3}F_{\frac{G_{u+1}\sigma^{(u)}}{\kappa}}(z_3-\frac{G_{u+1}}{\kappa}t_{k,p}^{(u)})\\
        & \equiv
        \sum_{k' \in \mathbb{Z}} b_{k',p}^{(u+1)} F_{\sigma_{u+1}}(\xi'_p-t_{k',p}^{(u+1)}).
    \end{split}
\end{align}
\end{widetext}
Here we indeed see that the PDFs of both quadratures on the $(u+1)$-th layer of error correction are composed of a sum of Gaussian functions. The new set of Gaussian functions have an identical STD $\sigma_{u+1}={\kappa\sigma_u}/{G_{u+1}}$, but different means $\{t_{k\prime, q}^{(u+1)}\},\;\{t_{k\prime, p}^{(u+1)}\}$ and weights $\{b_{k\prime, q}^{(u+1)}\},\; \{b_{k\prime, p}^{(u+1)}\}$. We can reorder the Gaussian functions into the following series,
\begin{align}
    \label{eq:SqzRep-updating-rule}
    \begin{split}
        &
        b_{(k,\ell),q}^{(u+1)} = b_{k,q}^{(u)} \int_{(\ell-\frac{1}{2})\sqrt{2\pi}}^{(\ell+\frac{1}{2})\sqrt{2\pi}}\diff{z_3}F_{\frac{G_{u+1}\sigma^{(u)}}{\kappa}}(z_3-\frac{\kappa}{G_{u+1}}t_{k,q}^{(u)}),\\
        &
        t_{(k,\ell),q}^{(u+1)} = -\kappa\frac{\sigma_u}{\sigma^{(u)}}(\ell\sqrt{2 \pi}-\frac{\kappa}{G_{u+1}}t_{k,q}^{(u)}),\\
        &
        b_{(k,\ell),p}^{(u+1)}=b_{k,p}^{(u)} \int_{(\ell-\frac{1}{2})\sqrt{2\pi}}^{(\ell+\frac{1}{2})\sqrt{2\pi}}\diff{z_3}F_{\frac{G_{u+1}\sigma^{(u)}}{\kappa}}(z_3-\frac{G_{u+1}}{\kappa}t_{k,p}^{(u)}),\\
        &
        t_{(k,m),p}^{(u+1)}= \kappa \ell \sqrt{2 \pi},\\
        &
        (k,\ell) \Rightarrow k^\prime.
    \end{split}
\end{align}
Now that we have proven the sum of Gaussian function part, next, we show the symmetry of Eq.~\eqref{eq:symmetry} also holds at the $(u+1)$-th layer, which makes $Q^{(u+1)}(\xi'_q)$ and $P^{(u+1)}(\xi'_p)$ even functions. Note that the $(k, \ell)$ term and the $(-k, -\ell)$ term match based on the given assumptions of Eq.~\eqref{eq:symmetry}. To be explicit, $b_{(k,\ell),q|p}^{(u+1)}= b_{(-k,-\ell),q|p}^{(u+1)}$ and $ t_{(k,\ell),q|p}^{(u+1)}= -t_{(-k,-\ell),q|p}^{(u+1)}$ give rise to $b_{k\prime,q|p}^{(u+1)} = b_{k\prime,q|p}^{(u+1)}$ and $t_{k\prime,q|p}^{(u+1)} = t_{k\prime,q|p}^{(u+1)}$.  
To complete the proof, we only need to verify that all assumptions are true at the $u=0$-th round. In that case, we have
\begin{align*}
    Q^{(0)}(\xi_q)=F_{\sigma_0}(\xi_q),
    \\
    P^{(0)}(\xi_p)=F_{\sigma_0}(\xi_p),
\end{align*}
which are indeed compositions of Gaussian functions, and the main peak (the only peak) has zero means. Also both functions are even functions. Therefore, we have completed the proof. The update rule can be used to evaluate the overall effects of the error correction.

From the above argument, we may calculate the variance of $Q^{(u+1)}(\xi_q)$ and $P^{(u+1)}(\xi_p)$ in terms of $\sigma_u$, $\sigma^{(u)}$ and $G_{u+1}$ easily since they are zero-mean.
\begin{align}
    \label{eq:SqzRep-variance}
    \begin{split}
    {\rm Var}_{q}^{(u+1)} &= (\frac{\kappa \sigma_u}{G_{u+1}})^2 + \sum_n b_{n,q}^{(u)}\left(t_{n,q}^{(u)}\right)^2,
    \\
    {\rm Var}_{p}^{(u+1)} &= (\frac{\kappa \sigma_u}{G_{u+1}})^2 + \sum_n b_{n,p}^{(u)}\left(t_{n,p}^{(u)}\right)^2.
    \end{split}
\end{align}

We briefly summarize the above analyses in preparation for the further optimization of the code. The initial channel noises are assumed to be independent AWGN. Correlation of noises of two channels is created by the symplectic transform Eq.~\eqref{eq:SqzRep-updating-rule}. Considering the main peak Eq.~\eqref{eq:main_peak}, the corresponding MLE and minimal variance estimator are identical. Using this estimator and properly choosing $\kappa$ according to Eq.~\eqref{eq:kappa}, the PDFs of noises are found to be a sum of Gaussian functions with updating rules given by Eqs.~\eqref{eq:SqzRep-updating-rule}.  Overall, from the list of STDs $\{\sigma_u\}_{u=0}^{n-1}$ over the $n$ modes, we can obtain the overall variances $({\rm Var}_{q}^{(n)},{\rm Var}_{p}^{(n)})$ as a function of $\{\sigma_u\}_{u=0}^{n-1}$ and the squeezing levels $\bm G=\{G_u\}_{u=1}^{n-1}$.

Now we consider the different orders of the channels. Let assume that $\bm G$ are fixed. When $G_{u+1} \sigma_{u} \gg \sigma^{(u)}$, we have $\kappa \rightarrow \sigma^{(u)}/\sigma_u$. So, if $\sigma^{(u)}$ is small, the leading term $\sigma^{(u+1)} \rightarrow \sigma^{(u)}/G_{u+1}$. 
For $n = 2$, this means $\sigma^{(2)} \rightarrow \sigma_
1/G$. Thus we would like to take $\sigma_1<\sigma_2$ in the two mode case to minimize the output noise, which is confirmed numerically. In the three modes case, this process continues as $\sigma^{(3)} \rightarrow \sigma^{(2)}/G_2 \rightarrow \sigma_1/(G_1 G_2)$. So we prefer to take $\sigma_1$ as the least noisy one and $\sigma_2$ as the second least noisy one, because combining $\sigma_1$ and $\sigma_2$ would make $\sigma^{(2)}$ smaller. By induction, we expect that the inverse order works well in general.

Now we show that the QEC code above reduces the noise to a level of $\sigma^{u+1} \sim \prod_{k=0}^u \sigma_{u}$ asymptotically, when considering only the main peak. The leading-order term is given by ${\kappa \sigma_u}/{G_{u+1}}$ from Eq.~\eqref{eq:SqzRep-variance}. And from Eq.~\eqref{eq:SqzRep-updating-rule}, we may take ${G_{u+1}\sigma^{(u)}}/{\kappa} = c \sqrt{2 \pi}$ in which $c \ll 1$ so that the noise falls in the range $[-\sqrt{\frac{\pi}{2}},\sqrt{\frac{\pi}{2}}]$ most of time. Then we have $\sigma^{(u+1)} = {\kappa \sigma_u}/{G_{u+1}} \sim \sigma_u\sigma^{(u)}\sim \prod_{k=0}^u \sigma_{u}$.
However, the exact STD is given by the square root of Eq.~\eqref{eq:SqzRep-variance}, which involves side peaks besides the main peak. For a certain number of channels with known STDs $\{\sigma_u\}_{u=0}^{n-1}$, we are able to arrange the order of the channels and choose the squeezing levels $\bm G$ to get a minimum ${\rm Var}_{q}^{(u+1)}+{\rm Var}_{p}^{(u+1)}$. Let us define the geometric mean value $\bar{\sigma} = (\prod_{k=0}^{n-1} \sigma_{u})^{\frac{1}{n}}$ and also define the output STD $\sigma_L$ through ${\rm Var}_{q}^{(u+1)}+{\rm Var}_{p}^{(u+1)} = 2\sigma_L^2 $. It's expected that $\sigma_L \sim \bar{\sigma}^N$ when all of the noise STDs are small.

Fig.~\ref{fig:PermutationList_SqzRep3} and Fig.~\ref{fig:PermutationList_SqzRep4} show the numerical results of permutation plots for the three-mode case and the four-mode case respectively.

Similar to the TMS code with heterogeneous independent AWGN, we derive the asymptotic curves in the same way. In the case of $n = 2$, the PDFs in both channels are exactly Gaussian. Let the STD be $\sigma_1$ for the channel transferring data mode and $\sigma_2$ for ancilla mode. Therefore, Eqs.~\eqref{eq:sq-rep PDFs1} and \eqref{eq:sq-rep PDFs2} reduce to:
\begin{align}
    &
    Q(\xi'_q) = \sum_{n \in \mathbb{Z}} b_n F_{\sigma^{(2)}}(\xi'_q-\kappa \frac{\sigma_1}{\sigma_2} n \sqrt{2 \pi}),\\
    &
    P(\xi'_p) = \sum_{n \in \mathbb{Z}} b_n F_{\sigma^{(2)}}(\xi'_p-\kappa n \sqrt{2 \pi}),
\end{align}
where $\sigma^{(2)} = {\kappa \sigma_1}/{G}$ and $b_n = \int_{(n-\frac{1}{2})\sqrt{2 \pi}}^{(n+\frac{1}{2})\sqrt{2 \pi}} F_{{G \sigma_2}/{\kappa}}(x)\diff{x}$. 
From Eq.~\eqref{eq:kappa}, when $G \gg 1$, $\kappa \approx {\sigma_2}/{\sigma_1}$. Let's take the sum over $n = 0, \pm 1$ to obtain the asymptotic result. Since $b_{\pm 1} \approx  \rm {Erfc}\left({\sqrt{\pi}}/{2 G \sigma_1}\right)/2$, using our definition for the total output noise we have:
\begin{align}
    \label{eq:SqzRep-2-sigmaL}
    2\sigma_L^2 
    & \equiv \rm{Var}_p + \rm{Var}_q \\
    &
    = \frac{2 \sigma_1^2 \sigma_2^2}{G^2 \sigma_1^2} + 2\pi \left(1+\frac{\sigma_2^2}{\sigma_1^2}\right)\rm{Erfc}\left(\frac{\sqrt{\pi}}{2 G \sigma_1}\right).
\end{align}
Let $x = {1}/{G^2 \sigma_1^2}$, then we need to solve the minimization of the function
\begin{align}
        f(x) = \sigma_1^2\sigma_2^2x+2\pi \left(1+\frac{\sigma_2^2}{\sigma_1^2}\right) \;{\rm Erfc}\left(\frac{\sqrt{\pi x}}{2}\right).
\end{align}
We find the optimum $x^*$ by fixing the derivative to be $0$ and then get the approximate value by solving the equation iteratively.
\begin{align}
    \label{eq:SqzRep-optimum-x}
    x^* &= 
    \frac{4}{\pi}\ln \left[\frac{\sqrt{\pi}}{4\sigma_1^2\sigma_2^2/(1+\sigma_2^2/\sigma_1^2)\sqrt{x^*}} \right]\\
    & \approx
    \frac{4}{\pi}\ln \left[\frac{\pi^{3/2}}{4\sigma_1^2\sigma_2^2/(1+\sigma_2^2/\sigma_1^2)}\right].
\end{align}
Therefore, we have the approximate $\sigma_L^2$ and also the asymptotic curves for $\sigma_L = t \sigma_2$ as
\begin{align}
    \label{eq:SqzRep-2-asymptotic-curves}
    &
    \sigma_L^2 \approx \frac{4\sigma_1^2\sigma_2^2}{\pi} \ln \left[\frac{\pi^{3/2}}{4\sigma_1^2\sigma_2^2/(1+\sigma_2^2/\sigma_1^2)}\right],
    \\
    &
    t^2 =\frac{ 4\sigma_1^2}{\pi} \ln\left[\frac{\pi^{3/2}}{4\sigma_1^2\sigma_2^2/(1+\sigma_2^2/\sigma_1^2)}\right].
\end{align}

\section{Quantum fidelity for a squeezed-vacuum state under noise}
\label{App:Fidelity}

The fidelity between a pure state $\ket{\psi}$ and a general state $\hat{\rho}$ equals
$
\mathcal{F}(\ket{\psi},\hat{\rho})=\sqrt{\braket{\psi |\hat{\rho} | \psi}}
$.
Consider a general displacement channel
\be 
\Phi_{p(\cdot)} \left(\hat{\rho}\right)=\int d^2\bm x p(x_1,x_2) \hat{D}\left(x_1,x_2\right)\hat{\rho} \hat{D}^\dagger\left(x_1,x_2\right),
\label{phi_N_disp_general}
\ee 
where $p(\cdot)$ is the distribution of displacement. 
Then we can calculate the fidelity of the state $\ket{\psi}$ before and after the general displacement channel by
\begin{align}
&\mathcal{F}\left(\ket{\psi},\Phi_{p(\cdot)}\left(\ketbra{\psi}\right)\right)=
\nonumber
\\
&\left[\int d^2\bm x p(x_1,x_2)
    |\braket{\psi|\hat{D}\left(x_1,x_2\right)|\psi}|^2\right]^\frac{1}{2} .
\label{eq:fidelity}
\end{align}

We consider a squeezed-vacuum state $\ket{\psi}=\ket{r,0}$ where $r>0$. The squeezed vacuum has one of the quadrature noise variance suppressed to $ e^{-2r}=10^{-x/10}$ of the vacuum noise, where $x$ is the squeezing in dB. We will use the displacement operator $\hat{D}(\alpha)$ for complex variable $\alpha = (x_1+i x_2)/\sqrt{2}$ and the squeeze operator $\hat{S}(r)$. 
We can calculate the inner product
\begin{align}
     \braket{r,0|\hat{D}\left(x_1,x_2\right)|r,0} &=  \braket{0|\hat{S}^\dagger (r)\hat{D}(\alpha)\hat{S}(r)|0}
     \nonumber\\
     & = \braket{0|\hat{D}(\alpha \cosh{r}+\alpha^* \sinh{r})|0}
     \nonumber\\
     & = \exp{-\frac{1}{2}\abs{\alpha \cosh{r}+\alpha^* \sinh{r}}^2} \nonumber\\
     & = \exp{-\frac{1}{4}(x^2_1 e^{2r}+x^2_2e^{-2r})}.
     \label{eq:chi-function-squeezed-vacuum}
\end{align}
Then we get the fidelity of a squeezed-vacuum after any displacement channel in Eq.~\eqref{eq:fidelity} as
\begin{align}
    &\mathcal{F}^2\left(\ket{r,0},\Phi_{p(\cdot)}(\hat{\rho})\right)=
    \nonumber \\
    & \int d^2\bm x p(x_1,x_2) \exp\left[-\frac{1}{2}(x^2_1 e^{2r}+x^2_2e^{-2r})\right]. 
\end{align}

Without QEC, the additive noise channel is Gaussian, and $p(x_1,x_2)=F_{\sigma}\left(x_1\right) F_{\sigma}\left(x_2\right)$ as in Eq.~\eqref{phi_N_disp}. We can obtain
\be 
\mathcal{F}^2\left(\ket{r,0},\Phi_{\sigma^2}(\hat{\rho})\right)= \frac{1}{\sqrt{1+\sigma^2 e^{2r}}\sqrt{1+\sigma^2 e^{-2r}}},
\label{eq:fidelity-no-QEC}
\ee 
where we have used the following formula
\begin{align*}
    \int_{-\infty}^{\infty} d x F_{\sigma} & \left(x-\mu \right) \exp \left(-\frac{1}{2}x^2 t\right)\\
    &=\frac{1}{\sqrt{1+\sigma^2 t}}\exp\left[-\frac{\mu^2 t}{2\left(1+\sigma^2 t\right)}\right].
\end{align*}
This agrees with the result obtained from the general formula for single-mode Gaussian states in Ref.~\cite{Weedbrook_2012}. When we have two additive noise channels $\Phi_{\sigma_1^2}$ and $\Phi_{\sigma_2^2}$, we will choose the channel with a smaller noise to obtain the best fidelity.

\begin{figure}[t]
    \centering
    \includegraphics[width=0.3\textwidth]{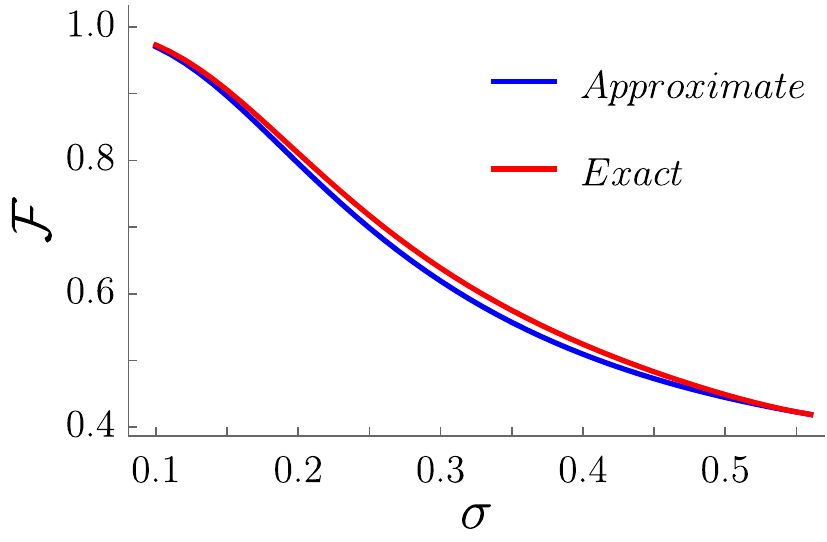}
    \caption{Comparison of Gaussian approximated fidelity and the exact result.}
    \label{fig:Fidelity-line}
\end{figure}

With QEC, the PDF of the additive noise is given by the sum of Gaussian functions in Eq.~\eqref{P_xi_general}. For simplicity, we consider the GKP-TMS code with a single GKP ancilla to demonstrate the effect of error correction. The distribution is given in Eq.~\eqref{sum_f}, and therefore the fidelity between the input and the QEC output $\Phi_{\rm QEC}^\prime(\hat{\rho})$ can be numerically calculated efficiently via
\begin{widetext}
\begin{align}
    \mathcal{F}^2\left(\ket{s,0},\Phi_{\rm QEC}^\prime(\hat{\rho})\right)&=\int d x_1 d x_2 \sum_{n,m\in \mathcal{Z}} b_n F_{{\sigma^{(2)}}}\left(x_1-\mu_n\right) b_m F_{{\sigma^{(2)}}}\left(x_2-\mu_m\right) \exp\left[-\frac{1}{2}(x^2_1 e^{2r}+x^2_2e^{-2r})\right] \nonumber \\
    & = \frac{1}{\sqrt{1+{\sigma^{(2)}}^2 e^{2r}}\sqrt{1+{\sigma^{(2)}}^2 e^{-2r}}} \sum_{n,m \in \mathcal{Z}}b_n b_m \exp\left[-\frac{\mu_n^2 t}{2\left(1+{\sigma^{(2)}}^2 e^{2r}\right)}\right] \exp\left[-\frac{\mu_m^2 t}{2\left(1+{\sigma^{(2)}}^2 e^{-2r}\right)}\right],
\label{eq:fidelity-QEC}
\end{align}
\end{widetext}
where $\sigma^{(2)}=\sigma_1\sigma_2/\sigma_G$ defined in the main paper.

It is natural to consider an approximation where we pretend the noise after the QEC is Gaussian and apply the Gaussian formula in Eq.~\eqref{eq:fidelity-no-QEC} to obtain
\be 
\mathcal{F}^2\left(\ket{s,0},\Phi_{\rm QEC}^\prime(\hat{\rho})\right)= \frac{1}{\sqrt{1+\sigma_L^2 e^{2r}}\sqrt{1+\sigma_L^2 e^{-2r}}}.
\label{eq:fidelity-QEC_approx}
\ee 
We compare the Gaussian approximated fidelity in Eq.~\eqref{eq:fidelity-QEC_approx} and the exact result in Eq.~\eqref{eq:fidelity-QEC} in Fig.~\ref{fig:Fidelity-line} for the case of $\sigma_1=\sigma_2=\sigma$, where a good approximation can be found. The exact result (red) in fact gives a higher fidelity than the Gaussian approximated result (blue).

\end{document}